\documentclass[prd,aps,12pt,preprintnumbers,amssymb,nofootinbib]{revtex4}
\usepackage{axodraw4j}
\usepackage{pstricks}
\usepackage{color}
\usepackage{amsmath}
\usepackage{slashed}
\usepackage{graphicx}
\usepackage{subfigure}
\def\be{\begin{equation}}
\def\ee{\end{equation}}
\def\bea{\begin{eqnarray}}
\def\eea{\end{eqnarray}}

\def\za{\alpha}
\def\zb{\beta}

\def\lsim{\mathrel{\raise.3ex\hbox{$<$\kern-.75em\lower1ex\hbox{$\sim$}}} }
\def\gsim{\mathrel{\raise.3ex\hbox{$>$\kern-.75em\lower1ex\hbox{$\sim$}}} }

\begin{document}
\preprint{\vbox{\hbox{NCU-HEP-k055}
\hbox{Oct 2012}\hbox{ed. Dec 2012}}}

\vspace*{1.5in}
\title{Comprehensive Analysis on Lepton Flavor Violating Higgs Boson to \mbox{\boldmath $\mu^\mp \tau^\pm$} Decay in Supersymmetry without \mbox{\boldmath $R$} Parity}

\author{\vspace*{.5in}\bf Abdesslam Arhrib }
\email{aarhrib@ictp.it}
\affiliation{Department of Mathematics, Faculty of Science and Techniques, B.P
  416 Tangier, Morocco}

\author{\bf Yifan Cheng and Otto C. W. Kong}
\email{otto@phy.ncu.edu.tw}
\affiliation{Department of Physics and 
Center for Mathematics and Theoretical Physics,
National Central University, Chung-li, Taiwan 32054
}
\begin{abstract}\vspace*{.5in}
In this paper we examine thoroughly the Higgs boson to $\mu^\mp \tau^\pm$ decay via
processes involving $R$ parity violating couplings. By means of full one-loop diagrammatic calculations, we found that even if known experimental constraints, particularly including the stringent sub-eV neutrino mass bounds, give strong restrictions on some of the $R$ parity violating parameters, the branching ratio could still achieve notable value in the admissible parameter space. Hence, the flavor violating leptonic decay is of interest to future experiments. We present here key results of our analysis. Based on the analysis, we give some comments on $h^0\rightarrow e^\mp \mu^\pm$ and $h^0\rightarrow e^\mp \tau^\pm$ also.
\end{abstract}
\maketitle

\newpage
\section{Introduction}
As we know, in the Standard Model (SM) the lepton number of each flavor is separately conserved. Thus lepton flavor violating (LFV) decays such as the Higgs boson to $\mu^\mp \tau^\pm$ are forbidden. However, neutrino oscillation experiments provide strong evidence that the lepton flavor conservation should be violated \cite{superk,K2K,SNO,KamLAND}. If lepton flavor violation can be observed in processes involving only SM particles, this would contribute an important probe to physics beyond the SM. Such processes, in particular the Higgs boson to $\mu^\mp \tau^\pm$ decay, deserve attention.

Looking into the literature, various sources or scenarios to accommodate LFV interactions have been introduced and analyzed. For example, adding heavy right-handed neutrinos can give neutrinos mixings and hence lepton flavor violation \cite{seesaw}. Also, a general two Higgs doublet model has LFV interactions due to Yukawa coupling matrices which can not be diagonalized simultaneously \cite{2hdm,2hdm2}. Under the framework of supersymmetry (SUSY), it is well known that
nonzero off-diagonal elements of soft SUSY breaking terms in the leptonic sector ($A^E$, $\tilde{m}^2_L$ and $\tilde{m}^2_E$, to be precisely defined below) generate LFV couplings. Moreover, the (total) lepton number itself may not be conserved. For the SUSY case, such $R$ parity violating (RPV) couplings also give interesting contributions to processes like the Higgs boson to $\mu^\mp \tau^\pm$ decay. 

While SUSY is undoubtedly a popular candidate theory for new physics, its existence so far lacks experimental evidence \cite{susyex}. Thus, some simple versions of the supersymmetric model, such as the constrained minimal supersymmetric standard model, have faced stringent challenges \cite{cmssm}. However, it has been pointed out that there is still room for the (minimal) supersymmetric standard model to accommodate existing experimental constraints \cite{susyalive,pmssm,nonuniversal,TeV}. For instance, the large mass spectrum for the majority of supersymmetric particles around or beyond 1 TeV has yet to be probed \cite{TeV}. The heavy spectrum is in accordance with the newly discovered boson mass $\cong$ 125 to 126 GeV \cite{ATLAS,CMS,125GeV}. A large portion of the parameter space remains uncovered in versions of the minimal supersymmetric standard model (MSSM) with more free parameters \cite{pmssm}. Nonuniversality of soft SUSY breaking masses is also a possible explanation for the nonobservation of supersymmetric signals \cite{nonuniversal}.

Under the scheme of the MSSM, various LFV decays such as $\tau\rightarrow\mu\gamma$, $\tau\rightarrow\mu X$, $\tau\rightarrow\mu\eta$, $\tau\rightarrow\mu\mu\mu$, and so on \cite{other}, as well as the Higgs boson to $\mu^\mp \tau^\pm$ decay \cite{soft,2hdm2} which we put our focus on in this paper, have been discussed. However, in many studies of the MSSM, $R$ parity is often imposed by hand to prevent proton decay and make the lightest supersymmetric particle a possible dark matter candidate. From the theoretical point of view, $R$ parity is {\it ad hoc} and not well motivated so long as the phenomenological (minimal) supersymmetric standard model is concerned \cite{Perez}. A generic supersymmetric standard model (without $R$ parity imposed), on the contrary, not only provides a convenient way to lepton flavor violation, but also has the advantage of a richer phenomenology including neutrino masses and mixings without introducing any extra superfield. Under the framework of SUSY with $R$ parity violation, there have been some studies \cite{rparity, ottoleptonic} on the issue of lepton flavor violation. Nevertheless, such studies were either limited to particular types of $R$ parity violation or did not take $h^0 \rightarrow \mu^\mp \tau^\pm$ into consideration. While recently both ATLAS and CMS \cite{ATLAS, CMS} reported discovery of a boson state which is essentially compatible with a SM-like Higgs, more data are needed to pin down its nature, and the flavor violating Higgs decay such as $h^0 \rightarrow \mu^\mp \tau^\pm$ is especially interesting at this moment. In this paper, we will investigate thoroughly the LFV Higgs boson to $\mu^\mp \tau^\pm$ decay from SUSY without $R$ parity via full diagrammatic calculations up to one-loop level. Under a reasonable choice of the experimentally viable parameter space, the most significant branching ratios of various RPV parameter combinations will be reported. Note that part of the key results has been reported, with limited presentation of analytical expressions and discussions, in a short letter \cite{1st}.

In following section, we summarize our basic formulation and parametrization of the generic supersymmetric standard model (without $R$ parity). Particularly, the neutral and charged Higgs mass terms, including loop corrections (up to two-loop for the neutral Higgs case), would be discussed. Then we give a sketch of our calculations and show numerical results from all possible RPV parameter combinations in section 3. Note that during our analysis we made no assumptions on the RPV parameters. The mass spectrum of all SUSY particles as well as the Higgs boson are kept within experimental constraints. Finally, we conclude this paper with some remarks in section 4. Lists of all one-loop diagrams and useful effective couplings will be given in the Appendices. We may be including more details than necessary,
particularly in the sense of showing some experimentally uninteresting results. We include those
to give a full picture about the physics involved, so that readers can appreciate the key
features leading to the interesting or uninteresting results. Some of the lessons one can learn
from the analysis would be useful for future studies of other related aspects of the model. Under the
same consideration, we give detailed expressions of the couplings involved and the Feynman diagrams 
in the Appendices. 
\section{Supersymmetric Standard Model without \mbox{\boldmath $R$} parity and Scalar Mass Matrices}
\subsection{Formulation and parametrization}
With the content of the minimal superfields spectrum, the most general renormalizable superpotential without $R$ parity can be written as
\begin{equation}
W=\epsilon_{ab}
\left[
\mu_\alpha \hat{H}^a_u \hat{L}^b_\alpha + h^u_{ik} \hat{Q}^a_i \hat{H}^b_u \hat{U}^C_k
+ \lambda^{'}_{\alpha jk} \hat{L}^a_\alpha \hat{Q}^b_j \hat{D}^C_k 
+ \frac{1}{2}\lambda_{\alpha\beta k}\hat{L}^a_\alpha \hat{L}^b_\beta \hat{E}^C_k
\right] 
+ \frac{1}{2}\lambda^{''}_{ijk}\hat{U}^C_i \hat{D}^C_j \hat{D}^C_k 
\end{equation}
where $(a,b)$ are SU(2) indices with $\epsilon_{12}=- \epsilon_{21}=1$, $(i,j,k)$ are the usual family (flavor) indices, and $(\alpha ,\beta)$ are extended flavor indices going from 0 to 3. Note that $\lambda$ is antisymmetric in the first two indices as required by SU(2) product rules while $\lambda^{''}$ is antisymmetric in the last two indices by $\text{SU(3)}_C$\,. The soft SUSY breaking terms can be written as follows: 
\begin{align}
V_{\rm soft} =&\,\epsilon_{ab}
  B_{\alpha} \,  H_{u}^a \tilde{L}_\alpha^b +
\epsilon_{ab} \left[ \,
  A^U_{ij} \tilde{Q}^a_i H^b_u \tilde{U}^\dagger_j
 +A^D_{ij} H^a_d \tilde{Q}^b_i \tilde{D}^\dagger_j 
 +A^E_{ij} H^a_d \tilde{L}^b_i \tilde{E}^\dagger_j \right]+\text{h.c.} \nonumber\\
&+\epsilon_{ab}\left[  
  A^{\lambda '}_{ijk} \tilde{L}^a_i \tilde{Q}^b_j \tilde{D}^\dagger_k 
 +\frac{1}{2} A^\lambda_{ijk} \tilde{L}^a_i \tilde{L}^b_j \tilde{E}^\dagger_k \right]
 +\frac{1}{2} A^{\lambda ''}_{ijk} \tilde{U}^\dagger_i \tilde{D}^\dagger_j \tilde{D}^\dagger_k+\text{h.c.} \nonumber \\
&+
 \tilde{Q}^\dagger \tilde{m}_Q^2 \,\tilde{Q} 
+\tilde{U}^{\dagger} 
\tilde{m}_U^2 \, \tilde{U} 
+\tilde{D}^{\dagger} \tilde{m}_D^2 
\, \tilde{D} 
+ \tilde{L}^\dagger \tilde{m}_L^2  \tilde{L}  
  +\tilde{E}^{\dagger} \tilde{m}_E^2 
\, \tilde{E}
+ \tilde{m}_{H_u}^2 \,
|H_{u}|^2 
\nonumber \\
& + \frac{M_1}{2} \tilde{B}\tilde{B}
   + \frac{M_2}{2} \tilde{W}\tilde{W}
   + \frac{M_3}{2} \tilde{g}\tilde{g}
+ \text{h.c.} \; ,
\end{align}
where $\tilde{L}^\dagger \tilde{m}^2_{\tilde{L}}\tilde{L}$ is given by a 4 $\times$ 4 matrix. $\tilde{m}^2_{L_{00}}$ corresponds to $\tilde{m}^2_{H_d}$ in MSSM, while $\tilde{m}^2_{L_{0k}}$'s give new mass mixings. Note that $\tilde{U}^\dagger$, $\tilde{D}^\dagger$, and $\tilde{E}^\dagger$ are the scalar components of the superfields $\hat{U}^C$, $\hat{D}^C$, and $\hat{E}^C$, respectively.

The above, together with the standard (gauged) kinetic terms, describe the full Lagrangian
of the model. We have four $\hat{L}$ superfields, which contain the components of the fermion doublet as $l^0$ and $l^-$, while their scalar partners are $\tilde{l}^0$ and $\tilde{l}^-$. In principle, the neutral scalar part $\tilde{l}^0_\alpha$ of all four $\hat{L}$ superfields can bear vacuum expectation values (VEVs). To make the analysis simple and the physics more transparent, we use a parametrization which picks a basis such that the direction of the VEV is singled out, i.e. only $\hat{L}_0$ bears a nonzero VEV among four $\hat{L}$'s. This procedure guarantees $\hat{L}_0$ can be always identified as $\hat{H}_d$ in MSSM. The two superfields have the same quantum number as the symmetry of the lepton number which makes the distinction between $\hat{L}$ and $\hat{H}_d$ by definition not part of the model. However, one should keep in mind that $\hat{H}_d$ may contain partly the charged lepton states. It is also worth mentioning here that the down quark and charged lepton Yukawa coupling matrix are both diagonal under our parametrization while the up quark Yukawa coupling is the product of Cabibbo-Kobayashi-Maskawa (CKM) factors and diagonal quark masses. The parametrization has the advantage 
that tree level RPV contributions to the neutral scalar mass matrix are described completely by the
$\mu_i$, $B_i$, and $\tilde{m}^2_{L_{0i}}$ parameters, which are well constrained to be small
even with just very conservative neutrino mass bounds imposed \cite{otto,otto98}. 

Now we turn to the issue about mass matrices of matter fields. In our framework, the three known charged leptons, together with two charginos, correspond to the mass eigenstates of a 5 $\times$ 5 charged fermion matrix ${\cal M}_C$, which can be diagonalized by two unitary matrices as $\mbox{\boldmath $V$}^\dagger{\cal M}_C\mbox{\boldmath $U$}=\text{diag}\left\{M_{\chi^-_n} \right\} \equiv \text{diag} \left\{M_{c1},M_{c2},m_e,m_\mu,m_\tau \right\}$.
For neutral fermions, we take four heavy neutralinos and three very light neutrinos as mass eigenstates under the scheme of a 7 $\times$ 7 neutral fermion mass matrix ${\cal M_N}$. By using a unitary matrix \mbox{\boldmath $X$}, the diagonalization can be done as $\mbox{\boldmath $X$}^T {\cal M_N}\mbox{\boldmath $X$}=\text{diag}\left\{M_{\chi^0_n}\right\} \equiv \text{diag}
\left\{M_{n_{i=1,4}},m_{\nu_1},m_{\nu_2},m_{\nu_3} \right\}$.
On considering the squark sectors, the up squark mass-squared matrix looks exactly the same as the one in MSSM, while the down squark one contains a new contribution from RPV terms. They can be diagonalized separately as 
${\cal D}^{u\dagger}{\cal M}^2_U{\cal D}^u = \rm{diag}\{{\cal M}^2_U\}$ and
${\cal D}^{d\dagger}{\cal M}^2_D{\cal D}^d = \rm{diag}\{{\cal M}^2_D\}$. All the mass matrices mentioned above can be found in \cite{otto}.
\subsection{Scalar mass matrices and loop corrections}
For the neutral scalar mass matrix, we have now five neutral complex scalar fields from $\hat{H_u}$ and four $\hat{L}_\alpha$'s. Explicitly, we write the $(1+4)$ complex fields in terms of their scalar and pseudoscalar parts, in the order \{$h^{0\dagger}_u$, $\tilde{l}^0_0$, $\tilde{l}^0_1$, $\tilde{l}^0_2$, $\tilde{l}^0_3$\} to form a full $10 \times 10$ (real and symmetric) mass-squared matrix, {which (in tree level) can be written as} 
\begin{equation}
{\cal M}_S^2 =
\left( \begin{array}{cc}
{\cal M}_{SS}^2 &
{\cal M}_{SP}^2 \\
({\cal M}_{SP}^{2})^{T} &
{\cal M}_{PP}^2
\end{array} \right) \; ,
\end{equation}
where the scalar, pseudoscalar, and mixing parts are
\begin{eqnarray}
{\cal M}_{SS}^2 &=&
\mbox{Re}({\cal M}_{{\phi}}^2)
+ 2\, {\cal M}_{{\phi\phi}}^2 \; ,
\nonumber \\
{\cal M}_{PP}^2 &=&
\mbox{Re}({\cal M}_{{\phi}}^2) \; ,
\nonumber \\
{\cal M}_{\! SP}^2 &=& - \mbox{Im}({\cal M}_{{\phi}}^2) \; ,
\end{eqnarray} 
respectively, with 
\begin{eqnarray}
{\cal M}_{{\phi\phi}}^2 =
\frac{1}{2} \, M_Z^2\,
\left( \begin{array}{ccc}
 \sin\!^2\! \beta  &  - \cos\!\beta \, \sin\! \beta
& \quad 0_{\scriptscriptstyle 1 \times 3} \\
 - \cos\!\beta \, \sin\! \beta & \cos\!^2\! \beta 
& \quad 0_{\scriptscriptstyle 1 \times 3} \\
0_{\scriptscriptstyle 3 \times 1} & 0_{\scriptscriptstyle 3 \times 1} 
& \quad 0_{\scriptscriptstyle 3 \times 3} 
\end{array} \right) \; ,
\end{eqnarray}
and
\begin{equation}
{\cal M}_{{\phi}}^2 =
\left( \begin{array}{cc}
\tilde{m}_{H_u}^2
+ \mu_{\za}^* \mu_{ \za}
+ M_{Z}^2\, \cos\!2 \beta 
\left[-\frac{1}{2}\right] 
& - (B_\za) \\
- (B_\za^*) &
\tilde{m}_{L}^2 
+ (\mu_{\za}^* \mu_{ \zb})
+ M_{Z}^2\, \cos\!2 \beta 
\left[ \frac{1}{2}\right]   I_{\scriptscriptstyle 4 \times 4} 
\end{array} \right) \; .
\end{equation}

 As for charged (colorless) scalars, we should treat charged Higgs and sleptons on an equal footing. The basis $\{ h^{+\dagger}_u, \tilde{l}^-_0, \tilde{l}^-_1, \tilde{l}^-_2, \tilde{l}^-_3, \tilde{l}^{+\dagger}_1, \tilde{l}^{+\dagger}_2, \tilde{l}^{+\dagger}_3 \}$ as $1+4+3$ form from $\hat{H_u}$, four $\hat{L}_\alpha$'s and three $\hat{E}^C_i$\,'s is used to write the $8 \times 8$ charged scalar mass-squared matrix, {which can be written as}
\begin{equation}
{\cal M}_E^2 =
\left( \begin{array}{ccc}
\widetilde{\cal M}_{H_u}^2 &
\widetilde{\cal M}_{LH}^{2\dag}  & 
\widetilde{\cal M}_{RH}^{2\dag}
\\
\widetilde{\cal M}_{LH}^2 & 
\widetilde{\cal M}_{LL}^{2} & 
\widetilde{\cal M}_{RL}^{2\dag} 
\\
\widetilde{\cal M}_{RH}^2 &
\widetilde{\cal M}_{RL}^{2} & 
\widetilde{\cal M}_{RR}^2  
\end{array} \right) \;,
\end{equation}
where
\begin{eqnarray}
\widetilde{\cal M}_{H_u}^2 &=&
\tilde{m}_{H_u}^2
+ \mu_{\za}^* \mu_{\za}
+ M_{Z}^2\, \cos\!2 \beta 
\left[ \,\frac{1}{2} - \sin\!^2\theta_{W}\right]
\nonumber \\
&\quad+&  M_{Z}^2\,  \sin\!^2 \beta \;
[1 - \sin\!^2 \theta_{W}]  \; ,
\nonumber \\
\widetilde{\cal M}_{LL}^2 &=&
\tilde{m}_{L}^2 +
m_{L}^\dag m_{L}
+ (\mu_{\za}^* \mu_{ \zb})
+ M_{Z}^2\, \cos\!2 \beta 
\left[ -\frac{1}{2} +  \sin\!^2 \theta_{W}\right] 
I_{\scriptscriptstyle 4 \times 4} \; 
\nonumber \\
&\quad+& \left( \begin{array}{cc}
 M_{Z}^2\,  \cos\!^2 \beta \;
[1 - \sin\!^2 \theta_{W}] 
& \quad 0_{\scriptscriptstyle 1 \times 3} \quad \\
0_{\scriptscriptstyle 3 \times 1} & 0_{\scriptscriptstyle 3 \times 3}  
\end{array} \right) \; ,
\nonumber \\
\widetilde{\cal M}_{RR}^2 &=&
\tilde{m}_{E}^2 +
m_{E} m_{E}^\dag
+ M_{Z}^2\, \cos\!2 \beta 
\left[  - \sin\!^2 \theta_{W}\right] 
I_{\scriptscriptstyle 3 \times 3}
\; ,
\end{eqnarray}
and
\begin{eqnarray}
\widetilde{\cal M}_{LH}^2
&=& (B_{\za}^*)  
+ \left( \begin{array}{c} 
\frac{1}{2} \,
M_Z^2\,  \sin\!2 \beta \;
[1 - \sin\!^2 \theta_{ W}]  \\
0_{\scriptscriptstyle 3 \times 1} 
\end{array} \right)
\; ,
\nonumber \\
\widetilde{\cal M}_{ RH}^2
&=&  -\,(\, \mu_i^*\lambda_{i0k}\, ) \; 
\frac{v_0}{\sqrt{2}} \; 
= (\, \mu_k^* \, m_k \, ) \hspace*{1in} \mbox{ (no sum over $k$)} \quad \; ,
\nonumber \\
(\widetilde{\cal M}_{RL}^{2})^{T} 
&=& \left(\begin{array}{c} 
0  \\   A^E 
\end{array}\right)
 \frac{v_0}{\sqrt{2}}
 - (\, \mu_{ \za}^*\lambda_{{ \za\zb}k}\, ) \; 
\frac{v_u}{\sqrt{2}}  \; ,
\end{eqnarray}
with {$m_{L}=
\mbox{diag}\{0,m_{E}\}= \mbox{diag}\{0,m_{1},
m_{2},m_{3}\}$. $m_i$'s ($\approx m_{e_i}$ under the small-$\mu_i$ scenario) are mass parameters in the charged fermion mass matrix \cite{otto}.} Furthermore, the two scalar mass-squared matrices can be diagonalized as
${{\cal D}^{s}}^{T}{\cal M}^{2}_{S}{\cal D}^{s}$ = diag\{${M^{2}_{S}}_{m=1,10}\}$ and
${{\cal D}^{l}}^{\dagger}{\cal M}^{2}_{E}{\cal D}^{l}$ = diag\{${M^{2}_{\tilde{\ell}}\!}_{n=1,8}\}$, which will become useful later. 

Different from MSSM, the physical scalar states are now a mixture of Higgs bosons and sleptons.
The RPV terms provide new contributions to the scalar mass matrices and hence the Higgs masses. In addition, radiative corrections, especially those from third generation quarks and squarks, could play an important role in the Higgs mass. Accordingly, we implement complete one-loop corrections \cite{Ellis} to matrix elements directly relating to Higgs bosons (CP-even, CP-odd and charged ones as well) during our computation. Moreover, the light Higgs mass should be treated delicately because of the newly discovered boson mass $\cong$ 125 to 126 GeV by the Large Hadron Collider (LHC) \cite{ATLAS,CMS}. Therefore we include further an estimation \cite{2loop} of key two-loop corrections in light Higgs related elements. \footnote{Though the Higgs bosons mix with the sleptons via RPV terms, we can still identify the Higgs bosons among other sleptons due to the foreseeable smallness of RPV parameters.} Note that radiative RPV corrections are typically too small to be taken into account; thus we study tree level RPV effects only.

Under the scheme of MSSM without $R$ parity, the one-loop effective Higgs potential is (recall that $\hat{H}_d\equiv \hat{L}_0$ after our parametrization is chosen) 
\begin{align}
V_{\rm eff}=
&\left(\tilde{m}^2_{H_u}+\left|\mu_\alpha\right|^2 \right)\left| H_u\right|^2
 +\left(\tilde{m}^2_{L_{00}}+\left|\mu_0\right|^2\right)\left| H_d\right|^2
 +\left(\epsilon_{ab}B_0 H^a_u H^b_d +\text{h.c.}\right) \nonumber\\
&+\frac{1}{8}\left(g_{\scriptscriptstyle 2}^2+g'^2\right)\left|H_u\right|^4  
+\frac{1}{8}\left(g_{\scriptscriptstyle 2}^2+g'^2\right)\left|H_d\right|^4 
 +\frac{1}{4}\left(g_{\scriptscriptstyle 2}^2-g'^2 \right)\left| H_u\right|^2\left| H_d\right|^2 
 -\frac{1}{2}g_{\scriptscriptstyle 2}^2 \left|\epsilon_{ab} H^a_u H^b_d \right|^2       \nonumber\\
&+\frac{3}{32\pi^2}\sum_{q=t,b}\left[\sum_{i=1,2}\tilde{m}^4_{q_i}
  \left(\ln\frac{\tilde{m}^2_{q_i}}{Q^2}-\frac{3}{2}\right)
  -2\bar{m}^4_q \left(\ln\frac{\bar{m}^2_q}{Q^2}-\frac{3}{2}\right) \right] \,,
\end{align}
where $Q$ is the renormalization scale which should be around the weak scale ($10^2$ to $10^3$ GeV). $\tilde{m}^2_{q_i}$ and $\bar{m}^2_q$ denote (Higgs background fields dependent) eigenvalues of the squark and quark mass matrices respectively.

By using the following linear expansion of Higgs bosons (with a relative complex phase for generality \footnote{It is basically a CP phase. In this study, the phase is set to be zero for simplicity.}),
\begin{equation}
H_{u}=\left(
\begin{array}{c}
h^{+}_{u} \\
\frac{1}{\sqrt{2}}\left(v_u+h^s_u-ih^a_u\right)
\end{array}\right) \, , \qquad
H_{d}=e^{i\theta_v}\left(
\begin{array}{c}
\frac{1}{\sqrt{2}}\left(v_d+h^s_d+ih^a_d\right) \\
h^{-}_{d}
\end{array}\right)\, , 
\end{equation}	
tadpole equations can be written as
\begin{align}
v_d \text{Re}(B_0 e^{i\theta_v})=
&\left(\tilde{m}^2_{H_u}+\left|\mu_\alpha\right|^2\right)v_u
 +\frac{1}{8}\left(g_{\scriptscriptstyle 2}^2+g'^2\right)v_u \left(v^2_u-v^2_d \right)  \nonumber\\  
&+\!\frac{3}{16\pi^2}\left[\sum_{q=t,b}\sum_{i=1,2}m^2_{\tilde{q}_i}
  \left\langle\frac{\partial \tilde{m}^2_{q_i}}{\partial h^s_u}\right\rangle\!
  \left(\ln\frac{m^2_{\tilde{q}_i}}{Q^2}-1\right)
  -2m^2_t \left\langle\frac{\partial \bar{m}^2_t}{\partial h^s_u}\right\rangle\!
  \left(\ln\frac{m^2_t}{Q^2}-1\right) \!\right] \nonumber\\
v_u \text{Re}(B_0 e^{i\theta_v})=
&\left(\tilde{m}^2_{L_{00}}+\left|\mu_0\right|^2\right) v_d
  +\frac{1}{8}\left(g_{\scriptscriptstyle 2}^2+g'^2\right)v_d\left(v^2_d -v^2_u \right)\nonumber\\ 
&+\!\frac{3}{16\pi^2}\left[\sum_{q=t,b}\sum_{i=1,2}m^2_{\tilde{q}_i}
  \left\langle\frac{\partial \tilde{m}^2_{q_i}}{\partial h^s_d}\right\rangle\!
  \left(\ln\frac{m^2_{\tilde{q}_i}}{Q^2}-1\right)
  -2m^2_b \left\langle\frac{\partial \bar{m}^2_b}{\partial h^s_d}\right\rangle\!
  \left(\ln\frac{m^2_b}{Q^2}-1\right) \!\right] \nonumber\\
v_{d(u)} \text{Im}(B_0 e^{i\theta_v})=
&+(-)\frac{3}{16\pi^2}\sum_{q=t,b}\sum_{i=1,2}m^2_{\tilde{q}_i}
  \left\langle\frac{\partial \tilde{m}^2_{q_i}}{\partial h^a_{u(d)}}\right\rangle
  \left(\ln\frac{m^2_{\tilde{q}_i}}{Q^2}-1\right) \,,
\end{align}
where $m^2_{\tilde{q}_i}=\left\langle \tilde{m}^2_{q_i} \right\rangle $ is the squark mass squared, while expressions for the derivatives with respect to the scalar fields in the bracket (including second derivatives used later) are complicated so we do not list them here. One can see \cite{Ellis} for example, for details. \footnote{There may be a sign difference between the expression for derivatives in the reference and ours due to the definition of linear expansion of scalars.}  

Tadpole equations along the direction of other scalars/sleptons can be obtained easily from scalar potential terms which are related to neutral sleptons: \footnote{Conceptually, $\tilde{l}_i^0$ is not the usual $\tilde{\nu}_i$ since $l_i^0$ deviates from $\nu_i$ slightly, with parameter $\mu_i$ characterizing the deviation between them. See \cite{otto} for details.}  
\begin{align}
V=&\sum_{i, j=1, 3}\Big[\big.
\left(\tilde{m}^2_{L_{ij}}+\mu^*_i \mu_j \right)\tilde{l}^{0*}_i \tilde{l}^0_j
 +\left(\tilde{m}^2_{L_{i0}}+\mu_0 \mu^*_i \right)h^0_d\tilde{l}^{0*}_i  
 +\left(\tilde{m}^2_{L_{0i}}+\mu^*_0 \mu_i \right)h^{0*}_d\tilde{l}^0_i \nonumber\\
&+\frac{1}{8}\left(g_{\scriptscriptstyle 2}^2+{g'}^2\right)\left(
                     \left|\tilde{l}^0_i\right|^2 \left|\tilde{l}^0_j\right|^2
                     -2\left|\tilde{l}^0_i\right|^2\left|h_u^0\right|^2
                     +2\left|\tilde{l}^0_i\right|^2\left|h_d^0\right|^2
                          \right)  
 -\left(B_i h^0_u \tilde{l}^0_i +\text{h.c.} \right)\big.\Big]  \,,                                     
\end{align}
while vanishing derivatives of $V$ give
\begin{equation}
B_i\tan\beta=\tilde{m}^2_{L_{0i}}+\mu^*_0\mu_i \,.
\end{equation}

{The exact form of tree level elements of scalar matrices are as mentioned above}, while the one-loop corrections from third generation quarks and squarks are
\begin{align}
{\cal M}^{\rm\scriptscriptstyle Loop}_{jk}=
&\,\frac{3}{16\pi^2}\sum_{q=t,b}\Big\{\big.\sum_{i=1,2} \left[
  \left\langle \frac{\partial \tilde{m}^2_{q_i}}{\partial \phi_j}\right\rangle
  \left\langle \frac{\partial \tilde{m}^2_{q_i}}{\partial \phi_k}\right\rangle
  \ln\frac{m^2_{\tilde{q}_i}}{Q^2}
 +m^2_{\tilde{q}_i}
  \left\langle\frac{\partial^2 \tilde{m}^2_{q_i}}{\partial \phi_j\partial \phi_k}\right\rangle
  \left(\ln\frac{m^2_{\tilde{q}_i}}{Q^2}-1\right)\right] \nonumber\\
&-2\left[
  \left\langle \frac{\partial \bar{m}^2_q}{\partial \phi_j}\right\rangle
  \left\langle \frac{\partial \bar{m}^2_q}{\partial \phi_k}\right\rangle
  \ln\frac{m^2_q}{Q^2}
 +m^2_q \left\langle\frac{\partial^2 \bar{m}^2_q}{\partial \phi_j\partial \phi_k}\right\rangle
  \left(\ln\frac{m^2_q}{Q^2}-1\right)\right] \big.\Big\} \;.
\end{align} 
In the case of neutral scalars, $j$ and $k$ can be any number among $1, 2, 6, 7$ which correspond to $h^s_u$, $h^s_d$, $h^a_u$ and $h^a_d$ respectively. As to the charged scalar case, $j$ and $k$ can only take the value of 1 or 2, with $\phi_j=\{h^+_u, h^+_d \}$ and $\phi_k=\{h^-_u, h^-_d\}$. By including the one-loop corrections mentioned above and the estimation of two-loop corrections \cite{2loop} to the scalar mass matrices, the numerical values of the Higgs masses can
be obtained with enough accuracy. 

\section{Calculations and Numerical Results}
At tree level, a neutral Higgs boson can decay into $\mu^-\tau^+$ or $\tau^-\mu^+$ directly via RPV neutral scalar-charged lepton-charged lepton coupling which is absent in MSSM. Otherwise, the neutral Higgs boson can decay through one-loop diagrams (or higher loop diagrams).
 We list in Appendix A all possible one-loop diagrams containing RPV couplings for a neutral scalar decaying to $\mu^-\tau^+$. The RPV effective couplings we used among all relevant mass eigenstates are listed in Appendix B. In our analysis, we diagonalize all the mass matrices numerically and deal directly with the mass eigenstates. The one-loop and two-loop corrections (as mentioned in section 2) to matrix elements which are most relevant to the Higgs mass are also implemented. We have fully calculated the decay amplitude of all (tree and one-loop) diagrams that may contribute. By encoding the analytical formulas of decay amplitude into the program, and using the \textit{LoopTools} \cite{looptools} program for the evaluation of loop functions, the numerical value of total amplitude and hence decay rate of $h^0 \rightarrow \mu^\mp \tau^\pm$ can be obtained. We include all the widths of significant decay channels in MSSM such as Higgs boson to $\bar{b}b$, $\tau^-\tau^+$, $WW^*$, $ZZ^*$, $\gamma\gamma$, and $gg$, plus the RPV decay rate of $h^0 \rightarrow \mu^\mp \tau^\pm$, to get the total width of Higgs decay. The branching ratio of $h^0 \rightarrow \mu^\mp \tau^\pm$ can then be obtained.

Our aim is to use a concrete setting that is compatible with known
constraints but not otherwise too restrictive, to illustrate what we expect to be
more generic features of the RPV signature. After considering the uncertainties in the experimental Higgs mass and loop corrections to Higgs mass terms, we kept the numerical light Higgs mass to be in the range of 123 to 127 GeV. Furthermore,
we adopt the relation $M_2=\frac{1}{3.5}M_3=2M_1$ between 
three gaugino masses and the condition that squarks of the first two families 
cannot be lighter than about $0.8 M_3$. Therefore we take soft 
SUSY breaking scalar masses 
$\tilde{m}^2_Q=\tilde{m}^2_U=\tilde{m}^2_D=(0.8 M_3\times \text{identity matrix})^2$ 
for simplicity in our analysis. The parameter setting is in accordance with the
gravity-mediated SUSY breaking picture \cite{Martin}, for instance. The other restrictions and assumptions we used can be found in Table 1.

\begin{center}
		\begin{tabular}{c@{\hspace{20pt}}c}
		\multicolumn{2}{l}{Table 1. List of the parameter ranges and conditions we adopted} \\
		\hline\hline
		Free parameters & Range \\
		\hline
           $\left|\mu_0\right|$, $M_2$, $\left|A_u\right|$, $\left|A_d\right|$ 
             and $\left|A^\lambda\right|$ & $\leq\text{2500 GeV}$\\
           $A_e$ &  zero, since its influence is negligible \\
                      $\tan\beta$ & 3 to 60  \\
           $\tilde{m}^2_E=\tilde{m}^2_L$ (without zeroth component) & $\leq(\text{2500 GeV})^2$ with off-diagonal elements zero\\
           $\tilde{m}^2_{L_{00}}$ & Constrained only by mass eigenvalues below\\
    \hline\hline
		Mass eigenvalues output & Range \\
		\hline
           Light Higgs mass & 123 to 127 GeV \\
           Heavy Higgs/sneutrino masses& 200 GeV to 3 TeV\\
           Charged Higgs/slepton masses & 200 GeV to 3 TeV\\[2pt]
		\hline
		\end{tabular} \\ \end{center}
		\medskip	\medskip

There are many different sources (e.g., flavor violating charged lepton decays like $\tau^-\to\mu^-e^+e^-$ \cite{ottomu}, leptonic radiative decays like $\mu\to e\gamma$ \cite{ottoleptonic}, semileptonic decays like $D^+\to \bar{K}^0l^+_i\nu_i$ \cite{Barbier},
experimental values of CKM matrix elements \cite{Barbier} and so on) which can give constraints on our 
RPV parameter setting. Among all the available constraints, the one from indirect evidence of the neutrino 
mass, i.e. $\sum_i m_{\nu_i}\lesssim\text{1 eV}$ \cite{neutrino} is quite crucial. 
Note that all LFV couplings/mass mixings that conserve $R$ parity have been turned off during our analysis. That is to single out the effects of the RPV ones. The reported numerical branching ratios are the most significant numbers we found under the framework. 

\subsection{Contribution from $B_i\,B_j$ combinations}

The constraints on this type of combination are mainly from neutrino mass experiments. The RPV parameter $B_i$ can give contributions to neutrino masses via one-loop diagrams \cite{Rakshit}. Generally speaking, larger sneutrino and neutralino masses will raise the upper bound of $B_i$.
				
Except for the combinations $B_2 B_3$ and $B_1 B_1$, $B_i B_j$ combinations can only give contributions to the decay from the Type2No.4 diagram (in Appendix A). Since heavy charged scalar masses will severely suppress this diagram, we can have relatively larger amplitudes only in the existence of light charged scalar(s). For the $B_2 B_3$ combination many diagrams contribute, hence its behavior is quite complicated. Basically, the decay amplitude from $B_2 B_3$ tends to increase when soft SUSY breaking scalar masses and gaugino masses get heavier due to the rise of the upper bound on $B_i$ from neutrino masses as mentioned above. Note that $B_2$ or $B_3$ alone can give contributions to the $h^0\rightarrow\mu^\mp \tau^\pm$ decay as well. Such contributions are, unavoidably, included in all combinations containing $B_2$ or $B_3$. Combinations that are not listed give zero contribution at one-loop level -- similar to the other kinds of RPV parameter combinations given below. Our results are shown in Table~2.

\begin{center}
		\begin{tabular}{c@{\hspace{20pt}}c@{\hspace{20pt}}c}
		\multicolumn{3}{l}{Table 2. $B_i \,B_j$ contributions to $Br(h^0\rightarrow\mu^\mp \tau^\pm)$} \\
		\hline\hline
				RPV parameter &\hspace*{1.5cm} & Admissible $Br$ within\\[-10pt]
		combinations &\hspace*{1.5cm} & known experimental constraints\\
						\hline
           $B_1 \,B_2$ & & $4\times10^{-22}$ \\
           $B_1 \,B_3$ & & $3\times10^{-22}$ \\
           $B_2 \,B_2$ & & $9\times10^{-23}$   \\
           $B_2 \,B_3$ & & $2\times10^{-11}$ \\
           $B_3 \,B_3$ & & $8\times10^{-23}$   \\[2pt]
		\hline
		\end{tabular} \\ \end{center}
		\medskip	\medskip

\subsection{Contribution from $B_i\,\mu_j$ combinations}

The $B_i\,\mu_j$ type of combination gets constrained from several sources. The values of $B_i$ and $B_i \mu_j$ are highly constrained separately by their loop contribution to neutrino masses \cite{Rakshit}. On the other hand, a nonzero $\mu_j$ will induce a tree level neutrino mass, hence it is also constrained. The nonobservation of leptonic radiative decays like $\mu \rightarrow e\gamma$, etc. also gives upper bounds on $B_i \mu_j$, say, $\left|B^*_1 \mu_3\right|$, $\left|B^*_2 \mu_3\right|$, $\left|B_3 \mu^*_1\right|$ and $\left|B_3 \mu^*_2\right| \lessapprox 10^{-4}\left|\mu_0\right|^3$; $\left|B^*_1 \mu_2\right|$ and $\left|B_2 \mu^*_1\right| \lessapprox 7\times10^{-7}\left|\mu_0\right|^3$
\cite{ottoleptonic}.  
 
    All $B_i \mu_j$ combinations except $B_2 \mu_3$ and $B_3 \mu_2$ can give contributions to the Higgs decay only from the Type2No.4 diagram. Again light charged scalars are preferred for the case. However, these contributions (from the Type2No.4 diagram) can not provide a significant branching ratio. Hence we have the uninterestingly tiny numbers as shown in Table~3.
  
    As for $B_2 \mu_3$ and $B_3 \mu_2$, both give contributions to the Higgs decay via many diagrams. Among them, the tree diagram (Fig. 1, left panel) is the most important over a wide range of parameter space. Especially for the $B_3 \mu_2$ combination, a key contribution to the decay amplitude is enhanced by the tau Yukawa coupling $y_{e_3}$ via a term
\[\approx{y_{e_3}M^*_2 B_3 \mu_2^*}\left(\tan\beta\sin\alpha-\cos\alpha\right)/[\sqrt{2}g_{\scriptscriptstyle 2}(\mu_0 M_2-M_W^2\sin2\beta)M^2_s]\]
($M^2_s$ denotes a generic real scalar mass eigenvalue). The latter makes the branching ratio from $B_3 \mu_2$ the largest among all $B_i \mu_j$'s. There is a similar feature for the contributions from the $B_2 \mu_3$ combination, but with a muon Yukawa $y_{e_2}$ instead. These two combinations get their most significant values under small $\mu_0$ and $M^2_s$ as can be seen from the expression above. Note that the contribution from loop diagrams is in general roughly smaller than that from the tree diagram, but can still be sizeable.

        \begin{center}
		\begin{tabular}{c@{\hspace{20pt}}c@{\hspace{20pt}}c}
		\multicolumn{3}{l}{Table 3. $B_i \,\mu_j$ contributions to $Br(h^0\rightarrow\mu^\mp \tau^\pm)$} \\
		\hline\hline
				RPV parameter &\hspace*{1.5cm} & Admissible $Br$ within\\[-10pt]
		combinations &\hspace*{1.5cm} & known experimental constraints\\
										\hline
           $B_1 \,\mu_2 $ & & $1\times10^{-24}$\\  
           $B_1 \,\mu_3 $ & & $1\times10^{-24}$  \\  

           $B_2 \,\mu_1 $ & & $9\times10^{-23}$  \\ 
           $B_2 \,\mu_2 $ & & $4\times10^{-26}$ \\            
           $B_2 \,\mu_3 $ & & $1\times10^{-15}$ \\ 

	   $B_3 \,\mu_1 $ & & $8\times10^{-23}$\\
	   $B_3 \,\mu_2 $ & & $1\times10^{-13}$  \\ 
	   $B_3 \,\mu_3 $ & & $4\times10^{-26}$  \\[2pt]
	   		\hline
		\end{tabular}\\ \end{center} \medskip\medskip

In fact, analyses similar to the above can be applied to $h^0\rightarrow e^\mp \mu^\pm$ and $h^0\rightarrow e^\mp \tau^\pm$ as well. The $B_i\mu_j$ contributions to $h^0\rightarrow e^\mp \mu^\pm$ are expected to be tiny due to the smallness of the corresponding Yukawa couplings $y_{e_1}$ and $y_{e_2}$. On the other hand, while the contributions from $B_1\mu_3$ are also suppressed by a relative factor of $y_{e_1}/y_{e_2}$, the contributions from $B_3\mu_1$ to $h^0\rightarrow e^\mp \tau^\pm$ could be roughly the same order as that of $h^0\rightarrow\mu^\mp \tau^\pm$. Hence, the $h^0\rightarrow e^\mp \tau^\pm$ decay may also be of interest. 

\subsection{Contribution from $B_i\,\lambda$ combinations}

Apart from the constraint on the $B_i$ parameters, the $\lambda$ type parameters are bounded by charged current experiments \cite{Barbier}. Generally speaking, increasing soft SUSY breaking slepton masses and gaugino masses leads to heavier charged slepton, sneutrino and neutralino masses and hence raises the upper bounds for $B_i$ and $\lambda$. 

\begin{center}
		\begin{tabular}{c@{\hspace{20pt}}c@{\hspace{20pt}}c}
		\multicolumn{3}{l}{Table 4. $B_i\,\lambda$ contributions to $Br(h^0\rightarrow\mu^\mp \tau^\pm)$} \\
		\hline\hline
				RPV parameter &\hspace*{1.5cm} & Admissible $Br$ within\\[-10pt]
		combinations &\hspace*{1.5cm} & known experimental constraints\\
										\hline
               $B_1 \,\lambda_{123}$ & & $1\times10^{-5\hspace{4pt}}$  \\
               $B_1 \,\lambda_{132}$ & & $3\times10^{-5\hspace{4pt}}$  \\
               $B_1 \,\lambda_{232}$ & & $4\times10^{-22}$  \\
               $B_1 \,\lambda_{233}$ & & $5\times10^{-25}$  \\
               $B_2 \,\lambda_{123}$ & & $7\times10^{-23}$ \\
               $B_2 \,\lambda_{131}$ & & $9\times10^{-24}$  \\
               $B_2 \,\lambda_{132}$ & & $5\times10^{-22}$  \\
               $B_2 \,\lambda_{232}$ & & $3\times10^{-5\hspace{4pt}}$  \\
               $B_2 \,\lambda_{233}$ & & $7\times10^{-23}$  \\
               $B_3 \,\lambda_{121}$ & & $5\times10^{-24}$  \\
               $B_3 \,\lambda_{123}$ & & $7\times10^{-23}$  \\
               $B_3 \,\lambda_{132}$ & & $5\times10^{-22}$  \\
               $B_3 \,\lambda_{232}$ & & $5\times10^{-22}$ \\
               $B_3 \,\lambda_{233}$ & & $3\times10^{-5\hspace{4pt}}$  \\[2pt]
		\hline
		\end{tabular}\\ \end{center}
\medskip\medskip

Among all the $B_i \lambda$ combinations, $B_1 \lambda_{123}$, $B_1 \lambda_{132}$, $B_2 \lambda_{232}$ and $B_3 \lambda_{233}$ are most important. They can provide large amplitudes via tree level diagrams (Fig. 1, middle panel), which are roughly 1 order of magnitude larger than that from loop diagrams. The amplitude can be approximated by ${\cal M}\approx{B_i\lambda}(\tan\beta\sin\alpha-\cos\alpha)/\left(\sqrt{2}M^2_s\right)$, where $\alpha$ is the mixing angle between two CP-even neutral Higgs bosons. Even though heavy sneutrino masses tend to suppress the amplitudes, they would relax the bounds on $B_i$ and $\lambda$ more significantly, and hence are favorable (Fig.~2, left panel).
Moreover, $\mu_0$ should not be too small in order to make the product of $B_i$ and $\lambda$ be below the bounds from leptonic radiative decays, i.e. $\left|B^*_1 \lambda_{132}\right|$, 
$\left|B_1 \lambda^*_{123}\right|$, $\left|B^*_2 \lambda_{232}\right|$ 
and $\left|B_3 \lambda^*_{233}\right| \lessapprox 1.4\times10^{-3}\left|\mu_0\right|^2$ 
\cite{ottoleptonic}. It is noteworthy that even under the stringent neutrino mass $\lesssim$ 1 eV constraint, the four combinations could give branching ratios beyond $10^{-5}$ (Fig.~2, right panel), which may be large enough to be probed at the LHC (or future linear collider). 
As to other $B_i \lambda$ combinations, they  
can be from several diagrams. However, they only play minor roles and hardly give any meaningful branching ratio, as shown in Table~4. 

As a matter of fact, the class of $B_i \lambda$ combinations gives the most important contributions to the flavor violating Higgs decays among all RPV parameter combinations. Moreover, the approximation of tree level amplitudes as above could apply to $h^0 \to e^\mp \tau^\pm$ and $h^0 \to e^\mp \mu^\pm$ as well. As a result, under the same parameter setting, it is expected for $h^0 \to e^\mp \tau^\pm$ and $h^0 \to e^\mp \mu^\pm$ to give branching ratios with roughly the same order of magnitude as in $h^0 \to \mu^\mp \tau^\pm$. However, it has been pointed out \cite{Blankenburg} that the LFV effective coupling between a light Higgs boson, electron and muon could not be large because of the constraint set by two-loop Barr-Zee diagrams \cite{BarrZee} on $\mu \to e\gamma$. Therefore, only $h^0 \to e^\mp \tau^\pm$ is expected to give a branching ratio comparable to that of $h^0 \to \mu^\mp \tau^\pm$.    		
		
\subsection{Contribution from $B_i\,A^\lambda$ combinations}

Under our parametrization, $A^\lambda$'s do not contribute to radiative decays such as $b \to s\gamma$ in one-loop level \cite{ottoleptonic}. Therefore, $A^\lambda$'s do not have known experimental constraints and, naively, can take any value. But the $B_i$ parameters are limited by loop neutrino masses as before. Contributions from $B_i A^\lambda$ may be quite interesting since this will be like the first experimental signature of the RPV $A$ parameters. However, an $A^\lambda$ only plays its role in the Higgs decay through Type2No.4 diagram with a neutral scalar-charged scalar-charged scalar ($h^0\phi^+\phi^-$) coupling. It is then expected to give a larger contribution at low charged scalar mass (Fig. 3, left panel).
 
\begin{center}
		\begin{tabular}{c@{\hspace{20pt}}c@{\hspace{20pt}}c}
		\multicolumn{3}{l}{Table 5. $B_i\,A^\lambda$ contributions to $Br(h^0\rightarrow\mu^\mp \tau^\pm)$} \\
		\hline\hline
				RPV parameter &\hspace*{1.5cm} & Admissible $Br$ within\\[-10pt]
		combinations &\hspace*{1.5cm} & known experimental constraints\\
										\hline
           $B_1 \,A^\lambda_{123}$ & & $5\times10^{-11}$ \\
           $B_1 \,A^\lambda_{132}$ & & $5\times10^{-11}$ \\
           $B_2 \,A^\lambda_{232}$ & & $5\times10^{-11}$  \\
           $B_3 \,A^\lambda_{233}$ & & $5\times10^{-11}$  \\[2pt]
		\hline
		\end{tabular}\\ \end{center}
\medskip\medskip

In our parameter setting, branching ratios from $B_i A^\lambda$ combinations can reach the order of $10^{-11}$ at most as shown in Table~5. However, if we allow $A^\lambda$ to be larger than hundreds of TeV, notable branching ratios are possible. Since decay rate is proportional to amplitude squared and hence $A^\lambda$ squared, it is easy to see how the branching ratio changes as $A^\lambda$ increases. As an example, we illustrate in Fig. 3 (both left and right panels) the branching ratio from the $B_2 A^\lambda_{232}$ contribution for $A^\lambda_{232}=2500$ GeV and 2500 TeV. In the extreme case of $A^\lambda_{232}=2500$ TeV, the branching ratio could reach the order of $10^{-5}$.

\subsection{Contribution from $\mu_i\,\lambda$ combinations}

All $\mu_i\,\lambda$ combinations which can contribute to $h^0\rightarrow\mu^\mp \tau^\pm$ at one-loop level, except $\mu_1 \lambda_{123}$ and $\mu_1 \lambda_{132}$, are constrained by their loop contributions to neutrino masses \cite{Rakshit}. Again, a single $\mu_i$ is constrained by its contribution to the tree level neutrino mass. Leptonic radiative decays also give upper bounds on $\mu_i\,\lambda$, i.e. $\left|\mu^*_2 \lambda_{232}\right|, \left|\mu^*_1 \lambda_{132}\right|, \left|\mu_3 \lambda^*_{233}\right|$ and $\left|\mu_1 \lambda^*_{123}\right| \lessapprox 7.0\times10^{-4}\left|\mu_0\right|$ \cite{ottoleptonic}. Further bounds for single $\lambda$ by charged current experiments can be found in \cite{Barbier}.

Many diagrams contribute to the $h^0\rightarrow\mu^\mp \tau^\pm$ process via $\mu_i\,\lambda$ combinations. Among these, Type1No.3 and Type1No.4 diagrams play the most important roles. The requirement of neutrino mass $\lesssim$ 1 eV still sets the most stringent bounds as in the case of other type combinations. However, $\mu_1 \lambda_{123}$ and $\mu_1 \lambda_{132}$ do not give loop contribution to neutrino masses, and thus they are mainly bounded by the constraints from radiative leptonic decays. Generally speaking, large slepton masses are favorable in order to have larger branching ratios since they can relax the constraints from loop neutrino masses and raise the upper bounds on the $\lambda$'s.

\begin{center}     
		\begin{tabular}{c@{\hspace{20pt}}c@{\hspace{20pt}}c}
		\multicolumn{3}{l}{Table 6. $\mu_i\,\lambda$ contributions to $Br(h^0\rightarrow\mu^\mp \tau^\pm)$} \\
		\hline\hline
				RPV parameter &\hspace*{1.5cm} & Admissible $Br$ within\\[-10pt]
		combinations &\hspace*{1.5cm} & known experimental constraints\\
										\hline
                 $\mu_1 \,\lambda_{123}$ & & $5\times10^{-8\hspace{4pt}}$ \\
                 $\mu_1 \,\lambda_{132}$ & & $5\times10^{-8\hspace{4pt}}$ \\ 
                 $\mu_2 \,\lambda_{232}$ & & $3\times10^{-12}$ \\
                 $\mu_2 \,\lambda_{131}$ & & $2\times10^{-24}$ \\
                 $\mu_3 \,\lambda_{233}$ & & $1\times10^{-14}$ \\ 
                 $\mu_3 \,\lambda_{121}$ & & $1\times10^{-24}$ \\[2pt]
		\hline
		\end{tabular}\\ \end{center}
\medskip		\medskip

In any case, branching ratios from $\mu_i\,\lambda$ can only achieve at most the order of $10^{-8}$ in our analysis because of the stringent constraints from leptonic decays. Our results are shown in Table~6.  
		
\subsection{Contribution from the other insignificant combinations}

In addition to the above combinations, there are some other types of combinations (i.e., $B_i\,\lambda'$, $\mu_i\,\mu_j$, $\mu_i\,\lambda'$, $\lambda\,\lambda$ and $\lambda'\,\lambda'$) which can merely give negligible contributions. Hence we only list the combinations which are most illustrative or give the largest branching ratios in each type of combination, as shown in Table~7. Note that the types of combinations which are not mentioned, $\lambda''\,\lambda''$ for example, give zero contributions at one-loop level.
		
In the $B_i\,\lambda'$ combinations, besides the constraints mentioned before on $B_i$, $\lambda'$ also gets constrained by charged/neutral current experiments \cite{Barbier,lambdaprime}. $B_i \lambda'$ combinations contribute to $h^0\rightarrow\mu^\mp \tau^\pm$ mainly via No.1 and No.2 diagrams of Type 6 and 7 (in Appendix A). To get better branching ratios, it is advantageous if we raise the upper bounds on $B_i$ by the heavy sneutrino and neutralino masses. Heavy squark masses could also raise the upper bounds on $\lambda'$. However, in our computation, contributions from $B_i \lambda'$ can not provide sizable branching ratios.  

As to $\mu_i \mu_j$ combinations, only $\mu_2\,\mu_3$ contributes to $h^0\rightarrow\mu^\mp \tau^\pm$ up to the one-loop level. With nonzero $\mu_i$\,, one of the neutrinos gets tree level mass. However, leptonic radiative
decays set more stringent bounds on $\mu_2\,\mu_3$ than the neutrino mass does \cite{ottomu}, i.e., 
\[\frac{\left|\mu_2\mu_3\right|}{\mu_0 M_2-M^2_W\sin2\beta} \lessapprox 4.3\times 10^{-3}(1+\tan^2\beta)\frac{\mu_0 M_2-M^2_W\sin2\beta}{M^2_W}\,.\]
Interestingly enough, though the $\mu_2 \mu_3$ combination contributes to the decay in tree level (Fig. 1, right panel), a loop contribution from Type1No.4 diagram is generally more important due to the smallness of neutrino masses in the loop. For example, where $\mu_2\mu_3$ gives its most significant branching ratio, the amplitude from loop diagrams compared to that from the tree diagram is roughly 10000:1. At any rate, $\mu_2 \mu_3$ could only give a negligible branching ratio. 
								
												\begin{center}
		\begin{tabular}{c@{\hspace{20pt}}c@{\hspace{20pt}}c}
		\multicolumn{3}{l}{Table\! 7.\! Most interesting examples in other RPV combinations} \\
		\hline\hline
				RPV parameter &\hspace*{1.5cm} & Admissible $Br$ within\\[-10pt]
		combinations &\hspace*{1.5cm} & known experimental constraints\\
										\hline
           $B_2 \,\lambda'_{333}$ & & $1\times10^{-14}$                   \\
           $B_3 \,\lambda'_{233}$ & & $1\times10^{-15}$        \\
        \hline
           $\mu_2\,\mu_3\hspace*{8pt}$ & & $2\times10^{-18}$  \\
        \hline
           $\mu_2\, \lambda'_{323}$ & & $3\times10^{-18}$   \\
           $\mu_3\, \lambda'_{223}$ & & $5\times10^{-19}$  \\
        \hline
           $\lambda_{232}\, \lambda_{233}$ & & $2\times10^{-19}$  \\
           $\lambda_{121}\, \lambda_{131}$ & & $1\times10^{-15}$ \\
           $\lambda_{123}\, \lambda_{133}$ & & $2\times10^{-10}$ \\
        \hline
           $\lambda'_{211}\, \lambda'_{311}$ & & $7\times10^{-18}$  \\
           $\lambda'_{222}\, \lambda'_{322}$ & & $4\times10^{-22}$  \\
           $\lambda'_{223}\, \lambda'_{323}$ & & $4\times10^{-12}$ \\
           $\lambda'_{233}\, \lambda'_{333}$ & & $5\times10^{-26}$  \\[2pt]
		\hline
		\end{tabular}\\ \end{center}  
\medskip	\medskip	

Among all $\mu_i\lambda'$\,'s which give nonzero contributions, some combinations
are constrained by their loop contributions to neutrino masses \cite{Rakshit}. Besides, every $\mu_i\,\lambda'$ is bounded by tree level neutrino mass constraints on $\mu_i$ and experimental constraints on single $\lambda'$ \cite{Barbier,lambdaprime}. In this type of combination, there is no obvious dominant diagram. Several diagrams can give comparable major contributions to the $h^0\rightarrow\mu^\mp \tau^\pm$ process. Generally speaking, heavy gaugino masses can relax the tree level neutrino mass constraints while heavy down squark masses can raise the upper bounds of $\lambda'$ and relax loop neutrino mass constraints; hence they are favorable for larger branching ratios. Unfortunately, in the whole parameter space, it is hard for the $\mu_i\lambda'$ to give any significant branching ratios. 		

Contributions from $\lambda\,\lambda$ combinations are mainly from No.3 and No.4 diagrams of Type 2, 6, and 7. Among all $\lambda\,\lambda$ combinations which contribute to $h^0\rightarrow\mu^\mp \tau^\pm$, only  $\lambda_{232}\lambda_{233}$ and $\lambda_{121}\lambda_{131}$ are constrained by their loop contributions to neutrino masses \cite{Rakshit}. However, leptonic decays could also provide upper bounds for $\lambda\lambda$ combinations \cite{ottoleptonic,Barbier,doublelambda}. Specifically, the neutrino mass constraint on $\lambda_{121}\, \lambda_{131}$ contains a factor of electron mass, and hence is relaxed by the smallness of electron mass. Therefore the branching ratio from $\lambda_{121}\, \lambda_{131}$ is mainly limited by restriction from leptonic decays. On the other hand, the neutrino mass constraint on $\lambda_{232}\, \lambda_{233}$ is enhanced by a $\tau$ mass factor. The latter gives a major restriction on the branching ratio from $\lambda_{232}\, \lambda_{233}$.
The most significant branching ratio from $\lambda\,\lambda$ we can have is of the order $10^{-10}$.

Many diagrams can contribute to the Higgs decay via $\lambda'\,\lambda'$ combinations. Among these, Type1No.2 and Type2No.1 diagrams are the most important ones. Just like the $\lambda\,\lambda$ case, among all $\lambda'\lambda'$ combinations which contribute to $h^0\rightarrow\mu^\mp \tau^\pm$, only $\lambda'_{211}\lambda'_{311}$, $\lambda'_{222}\lambda'_{322}$ and $\lambda'_{233}\lambda'_{333}$ as listed contribute to neutrino masses \cite{Rakshit} and hence get additional constraints. Besides, radiative $B$ decays and leptonic decays also give upper bounds on $\lambda'\,\lambda'$ \cite{bsr,Barbier,doublelambda,doublelprime}. Particularly,
constraints on $\lambda'_{211}\lambda'_{311}$, $\lambda'_{222}\lambda'_{322}$ and $\lambda'_{233}\lambda'_{333}$ are suppressed/enhanced separately by the electron, muon, and tau mass factors. This makes the differences between their branching ratios. 
Nevertheless, $\lambda'\,\lambda'$ type of combination could only give negligible contributions to $h^0\rightarrow\mu^\mp \tau^\pm$.
		
\section{Summary}
We have analyzed thoroughly Higgs to $\mu^\mp \tau^\pm$ decay in the framework of the minimal supersymmetric standard model without $R$ parity. By means of full one-loop diagrammatic calculations and taking the RPV terms as the only source of lepton flavor violation, we showed that the branching ratio of $h^0\rightarrow \mu^\mp \tau^\pm$ could exceed $10^{-5}$ without contradicting experimental constraints. We pull together the most interesting RPV parameter combinations and corresponding branching ratios in Table~8 for easy reference. The numbers in the parentheses indicate the branching ratios in the case of $A^\lambda =2500$ TeV as mentioned in the $B_i A^\lambda$ section. Moreover, $h^0\rightarrow e^\mp \tau^\pm$ is expected to be able to give roughly the same order of branching ratio with that of $h^0\rightarrow \mu^\mp \tau^\pm$ from RPV terms, while $h^0\rightarrow e^\mp \mu^\pm$ is suppressed due to stringent constraint from two-loop Barr-Zee diagrams.

    \begin{center}
		\begin{tabular}{c@{\hspace{20pt}}c@{\hspace{20pt}}c}
		\multicolumn{3}{l}{Table 8. Interesting contributions to branching ratio of $h^0\rightarrow\mu^\mp \tau^\pm$} \\
		\hline\hline
				RPV parameter &\hspace*{1.5cm} & Admissible $Br$ within\\[-10pt]
		combinations &\hspace*{1.5cm} & known experimental constraints\\
								\hline    
           $B_1 \,\lambda_{123}$ & & $1\times10^{-5\hspace{10pt}}$  \\
           $B_1 \,\lambda_{132}$ & & $3\times10^{-5\hspace{10pt}}$ \\
           $B_2 \,\lambda_{232}$ & & $3\times10^{-5\hspace{10pt}}$ \\
           $B_3 \,\lambda_{233}$ & & $3\times10^{-5}\hspace{10pt}$ \\
           $B_2 \,A^\lambda_{232}$ & & $\hspace{12pt}5\times10^{-11(-5)}$  \\
           $B_3 \,A^\lambda_{233}$ & & $\hspace{12pt}5\times10^{-11(-5)}$  \\[2pt]

	   		\hline
		\end{tabular}\\ \end{center} \medskip\medskip

Generally speaking, a heavy SUSY spectrum is preferred for large branching ratios of LFV Higgs decays obtainable from RPV couplings. The resulting relaxations of the experimental constraints from other processes and especially neutrino masses on the couplings leave more room for the Higgs decay. 
However, the statement may not hold for the contributions involving the $A^\lambda$ parameters. In the extreme case that such a parameter is larger than around hundreds of TeV, notable branching ratios are possible, especially with relatively light slepton masses (below 1 TeV). 
Meanwhile, a smaller value of the Higgs mass parameter $M_A$ is favored in the LFV Higgs decays of the RPV scenario.   

From an experimental point of view, a typical cross section of the MSSM 125
GeV Higgs boson at 8 TeV energy is of the order 10 pb. Short of a
  reliable full simulation study, we can only carry out a rough estimate on the
  observability of the $Br(h^0\to \mu^\mp \tau^\pm) \gtrsim
  10^{-5}$. With a luminosity of the order $10$ fb$^{-1}$, this would lead to
  several raw $\mu^\mp \tau^\pm$ events with almost no SM 
background. \footnote{Our estimate is likely to be on 
the optimistic side when detector properties are fully taken into 
consideration. Some complete experimental analyses with realistic 
cuts may be needed to improve the situation. The case for the 14TeV running
or a future linear collider will be much better. 
We also want to bring to the reader's attention that after 
we finished our work, a preprint \cite{simulation} on the relevant
branching ratio reach of the 8 TeV LHC appears, claiming a quite disappointing
number.} 
If we allow more free parameters or a larger parameter space during our analysis, the branching ratios can become even larger. Together with the 14 TeV energy for future LHC runs, we may have more events and a better chance to probe lepton flavor violation, and physics beyond the standard model.
		
\section*{Acknowledgements}
Y.C. and O.K. are partially supported by Research Grant No. NSC 99-
2112-M-008-003-MY3 of the National Science Council of Taiwan.

\newpage
\appendix
\section{One-Loop Feynman Diagrams in MSSM without \mbox{\boldmath $R$} Parity for the Neutral Higgs \mbox{\boldmath $\phi^0 \rightarrow \mu^-\tau^+$}}
\vspace*{1cm}
\begin{center}
  \begin{picture}(450,70)(20,-85)
    \SetWidth{0.5}
    \SetColor{Black}

    \Line[arrow,arrowpos=0.5,arrowlength=3,arrowwidth=1.1,arrowinset=0.2](80,-20)(50,-40)
    \Line[arrow,arrowpos=0.5,arrowlength=3,arrowwidth=1.1,arrowinset=0.2](50,-40)(80,-60)
    \Line[dash,dashsize=5,arrow,arrowpos=0.5,arrowlength=3,arrowwidth=1.1,arrowinset=0.2](80,-60)(80,-20)
    \Line[arrow,arrowpos=0.5,arrowlength=3,arrowwidth=1.1,arrowinset=0.2](110,-10)(80,-20)
    \Line[arrow,arrowpos=0.5,arrowlength=3,arrowwidth=1.1,arrowinset=0.2](80,-60)(110,-70)
    \Line[dash,dashsize=5](50,-40)(20,-40)
    \Text(90,-11)[lb]{\Black{$\tau$}}
    \Text(60,-23)[lb]{\Black{$d_{i}$}}
    \Text(60,-66)[lb]{\Black{$d_{j}$}}
    \Text(85,-43)[lb]{\Black{$\tilde{u}_{k}$}}
    \Text(90,-77)[lb]{\Black{$\mu$}}
    \Text(30,-55)[lb]{\Black{$\phi^{0}$}}
    \Vertex(80,-20){1.5}
    \Vertex(50,-40){1.5}
    \Vertex(80,-60){1.5}  
    \Text(65,-90)[]{\Black{Type1No.1}}
    
    \Line[arrow,arrowpos=0.5,arrowlength=3,arrowwidth=1.1,arrowinset=0.2](165,-40)(195,-20)
    \Line[arrow,arrowpos=0.5,arrowlength=3,arrowwidth=1.1,arrowinset=0.2](195,-60)(165,-40)
   \Line[dash,dashsize=5,arrow,arrowpos=0.5,arrowlength=3,arrowwidth=1.1,arrowinset=0.2](195,-20)(195,-60)
    \Line[arrow,arrowpos=0.5,arrowlength=3,arrowwidth=1.1,arrowinset=0.2](225,-10)(195,-20)
    \Line[arrow,arrowpos=0.5,arrowlength=3,arrowwidth=1.1,arrowinset=0.2](195,-60)(225,-70)
    \Line[dash,dashsize=5](165,-40)(135,-40)
    \Text(205,-11)[lb]{\Black{$\tau$}}
    \Text(175,-23)[lb]{\Black{$u_{i}$}}
    \Text(175,-64)[lb]{\Black{$u_{j}$}}
    \Text(200,-43)[lb]{\Black{$\tilde{d}_{k}$}}
    \Text(205,-77)[lb]{\Black{$\mu$}}
    \Text(145,-55)[lb]{\Black{$\phi^{0}$}}
    \Vertex(195,-20){1.5}
    \Vertex(165,-40){1.5}
    \Vertex(195,-60){1.5}
    \Text(180,-90)[]{\Black{Type1No.2}}
    
    \Line[arrow,arrowpos=0.5,arrowlength=3,arrowwidth=1.1,arrowinset=0.2](310,-20)(280,-40)
    \Line[arrow,arrowpos=0.5,arrowlength=3,arrowwidth=1.1,arrowinset=0.2](280,-40)(310,-60)
   \Line[dash,dashsize=5,arrow,arrowpos=0.5,arrowlength=3,arrowwidth=1.1,arrowinset=0.2](310,-20)(310,-60)
    \Line[arrow,arrowpos=0.5,arrowlength=3,arrowwidth=1.1,arrowinset=0.2](340,-10)(310,-20)
    \Line[arrow,arrowpos=0.5,arrowlength=3,arrowwidth=1.1,arrowinset=0.2](310,-60)(340,-70)
    \Line[dash,dashsize=5](280,-40)(250,-40)
    \Text(320,-11)[lb]{\Black{$\tau$}}
    \Text(290,-23)[lb]{\Black{$\chi^{-}_{i}$}}
    \Text(290,-69)[lb]{\Black{$\chi^{-}_{j}$}}
    \Text(315,-43)[lb]{\Black{$\phi^{0}_{k}$}}
    \Text(320,-77)[lb]{\Black{$\mu$}}
    \Text(260,-55)[lb]{\Black{$\phi^{0}$}}
    \Vertex(310,-20){1.5}
    \Vertex(280,-40){1.5}
    \Vertex(310,-60){1.5}
    \Text(295,-90)[]{\Black{Type1No.3}}
    
    \Line[arrow,arrowpos=0.5,arrowlength=3,arrowwidth=1.1,arrowinset=0.2](425,-20)(395,-40)
    \Line[arrow,arrowpos=0.5,arrowlength=3,arrowwidth=1.1,arrowinset=0.2](395,-40)(425,-60)
   \Line[dash,dashsize=5,arrow,arrowpos=0.5,arrowlength=3,arrowwidth=1.1,arrowinset=0.2](425,-20)(425,-60)
    \Line[arrow,arrowpos=0.5,arrowlength=3,arrowwidth=1.1,arrowinset=0.2](455,-10)(425,-20)
    \Line[arrow,arrowpos=0.5,arrowlength=3,arrowwidth=1.1,arrowinset=0.2](425,-60)(455,-70)
    \Line[dash,dashsize=5](395,-40)(365,-40)
    \Text(435,-11)[lb]{\Black{$\tau$}}
    \Text(405,-23)[lb]{\Black{$\chi^{0}_{i}$}}
    \Text(405,-69)[lb]{\Black{$\chi^{0}_{j}$}}
    \Text(430,-43)[lb]{\Black{$\phi^{-}_{k}$}}
    \Text(435,-77)[lb]{\Black{$\mu$}}
    \Text(375,-55)[lb]{\Black{$\phi^{0}$}}
    \Vertex(425,-20){1.5}
    \Vertex(395,-40){1.5}
    \Vertex(425,-60){1.5}
    \Text(410,-90)[]{\Black{Type1No.4}}

    \Line[dash,dashsize=5,arrow,arrowpos=0.5,arrowlength=3,arrowwidth=1.1,arrowinset=0.2](50,-160)(80,-140)
    \Line[dash,dashsize=5,arrow,arrowpos=0.5,arrowlength=3,arrowwidth=1.1,arrowinset=0.2](80,-180)(50,-160)
    \Line[arrow,arrowpos=0.5,arrowlength=3,arrowwidth=1.1,arrowinset=0.2](80,-140)(80,-180)
    \Line[arrow,arrowpos=0.5,arrowlength=3,arrowwidth=1.1,arrowinset=0.2](110,-130)(80,-140)
    \Line[arrow,arrowpos=0.5,arrowlength=3,arrowwidth=1.1,arrowinset=0.2](80,-180)(110,-190)
    \Line[dash,dashsize=5](50,-160)(20,-160)
    \Text(90,-131)[lb]{\Black{$\tau$}}
    \Text(60,-144)[lb]{\Black{$\tilde{u}_{i}$}}
    \Text(60,-187)[lb]{\Black{$\tilde{u}_{j}$}}
    \Text(85,-163)[lb]{\Black{$d_{k}$}}
    \Text(90,-198)[lb]{\Black{$\mu$}}
    \Text(30,-175)[lb]{\Black{$\phi^{0}$}}
    \Vertex(80,-140){1.5}
    \Vertex(50,-160){1.5}
    \Vertex(80,-180){1.5}  
    \Text(65,-210)[]{\Black{Type2No.1}}
    
    \Line[dash,dashsize=5,arrow,arrowpos=0.5,arrowlength=3,arrowwidth=1.1,arrowinset=0.2](195,-140)(165,-160)
    \Line[dash,dashsize=5,arrow,arrowpos=0.5,arrowlength=3,arrowwidth=1.1,arrowinset=0.2](165,-160)(195,-180)
   \Line[arrow,arrowpos=0.5,arrowlength=3,arrowwidth=1.1,arrowinset=0.2](195,-180)(195,-140)
    \Line[arrow,arrowpos=0.5,arrowlength=3,arrowwidth=1.1,arrowinset=0.2](225,-130)(195,-140)
    \Line[arrow,arrowpos=0.5,arrowlength=3,arrowwidth=1.1,arrowinset=0.2](195,-180)(225,-190)
    \Line[dash,dashsize=5](165,-160)(135,-160)
    \Text(205,-131)[lb]{\Black{$\tau$}}
    \Text(175,-144)[lb]{\Black{$\tilde{d}_{i}$}}
    \Text(175,-190)[lb]{\Black{$\tilde{d}_{j}$}}
    \Text(200,-163)[lb]{\Black{$u_{k}$}}
    \Text(205,-198)[lb]{\Black{$\mu$}}
    \Text(145,-175)[lb]{\Black{$\phi^{0}$}}
    \Vertex(195,-140){1.5}
    \Vertex(165,-160){1.5}
    \Vertex(195,-180){1.5}
    \Text(180,-210)[]{\Black{Type2No.2}}
    
    \Line[dash,dashsize=5,arrow,arrowpos=0.5,arrowlength=3,arrowwidth=1.1,arrowinset=0.2](310,-140)(280,-160)
    \Line[dash,dashsize=5,arrow,arrowpos=0.5,arrowlength=3,arrowwidth=1.1,arrowinset=0.2](280,-160)(310,-180)
   \Line[arrow,arrowpos=0.5,arrowlength=3,arrowwidth=1.1,arrowinset=0.2](310,-140)(310,-180)
    \Line[arrow,arrowpos=0.5,arrowlength=3,arrowwidth=1.1,arrowinset=0.2](340,-130)(310,-140)
    \Line[arrow,arrowpos=0.5,arrowlength=3,arrowwidth=1.1,arrowinset=0.2](310,-180)(340,-190)
    \Line[dash,dashsize=5](280,-160)(250,-160)
    \Text(320,-131)[lb]{\Black{$\tau$}}
    \Text(290,-144)[lb]{\Black{$\phi^{0}_{i}$}}
    \Text(289,-189)[lb]{\Black{$\phi^{0}_{j}$}}
    \Text(315,-163)[lb]{\Black{$\chi^{-}_{k}$}}
    \Text(320,-198)[lb]{\Black{$\mu$}}
    \Text(260,-175)[lb]{\Black{$\phi^{0}$}}
    \Vertex(310,-140){1.5}
    \Vertex(280,-160){1.5}
    \Vertex(310,-180){1.5}
    \Text(295,-210)[]{\Black{Type2No.3}}
   
    \Line[dash,dashsize=5,arrow,arrowpos=0.5,arrowlength=3,arrowwidth=1.1,arrowinset=0.2](425,-140)(395,-160)
    \Line[dash,dashsize=5,arrow,arrowpos=0.5,arrowlength=3,arrowwidth=1.1,arrowinset=0.2](395,-160)(425,-180)
   \Line[arrow,arrowpos=0.5,arrowlength=3,arrowwidth=1.1,arrowinset=0.2](425,-140)(425,-180)
    \Line[arrow,arrowpos=0.5,arrowlength=3,arrowwidth=1.1,arrowinset=0.2](455,-130)(425,-140)
    \Line[arrow,arrowpos=0.5,arrowlength=3,arrowwidth=1.1,arrowinset=0.2](425,-180)(455,-190)
    \Line[dash,dashsize=5](395,-160)(365,-160)
    \Text(435,-131)[lb]{\Black{$\tau$}}
    \Text(405,-144)[lb]{\Black{$\phi^{-}_{i}$}}
    \Text(403,-189)[lb]{\Black{$\phi^{-}_{j}$}}
    \Text(430,-163)[lb]{\Black{$\chi^{0}_{k}$}}
    \Text(435,-198)[lb]{\Black{$\mu$}}
    \Text(375,-175)[lb]{\Black{$\phi^{0}$}}
    \Vertex(425,-140){1.5}
    \Vertex(395,-160){1.5}
    \Vertex(425,-180){1.5}
    \Text(410,-210)[]{\Black{Type2No.4}}

    \Photon(80,-260)(50,-280){2}{4}
    \Line[dash,dashsize=5,arrow,arrowpos=0.5,arrowlength=3,arrowwidth=1.1,arrowinset=0.2](50,-280)(80,-300)
    \Line[arrow,arrowpos=0.5,arrowlength=3,arrowwidth=1.1,arrowinset=0.2](80,-260)(80,-300)
    \Line[arrow,arrowpos=0.5,arrowlength=3,arrowwidth=1.1,arrowinset=0.2](110,-250)(80,-260)
    \Line[arrow,arrowpos=0.5,arrowlength=3,arrowwidth=1.1,arrowinset=0.2](80,-300)(110,-310)
    \Line[dash,dashsize=5](50,-280)(20,-280)
    \Text(90,-251)[lb]{\Black{$\tau$}}
    \Text(58,-264)[lb]{\Black{$Z$}}
    \Text(58,-310)[lb]{\Black{$\phi^{0}_{j}$}}
    \Text(85,-283)[lb]{\Black{$\chi^{-}_k$}}
    \Text(90,-318)[lb]{\Black{$\mu$}}
    \Text(30,-295)[lb]{\Black{$\phi^{0}$}}
    \Vertex(80,-260){1.5}
    \Vertex(50,-280){1.5}
    \Vertex(80,-300){1.5}  
    \Text(65,-330)[]{\Black{Type3No.1}}   
         
    \Photon(195,-260)(165,-280){2}{4}
    \Line[dash,dashsize=5,arrow,arrowpos=0.5,arrowlength=3,arrowwidth=1.1,arrowinset=0.2](165,-280)(195,-300)
   \Line[arrow,arrowpos=0.5,arrowlength=3,arrowwidth=1.1,arrowinset=0.2](195,-260)(195,-300)
    \Line[arrow,arrowpos=0.5,arrowlength=3,arrowwidth=1.1,arrowinset=0.2](225,-250)(195,-260)
    \Line[arrow,arrowpos=0.5,arrowlength=3,arrowwidth=1.1,arrowinset=0.2](195,-300)(225,-310)
    \Line[dash,dashsize=5](165,-280)(135,-280)
    \Text(205,-251)[lb]{\Black{$\tau$}}
    \Text(173,-264)[lb]{\Black{$W$}}
    \Text(173,-310)[lb]{\Black{$\phi^{-}_{j}$}}
    \Text(200,-283)[lb]{\Black{$\chi^0_k$}}
    \Text(205,-318)[lb]{\Black{$\mu$}}
    \Text(145,-295)[lb]{\Black{$\phi^{0}$}}
    \Vertex(195,-260){1.5}
    \Vertex(165,-280){1.5}
    \Vertex(195,-300){1.5}
    \Text(180,-330)[]{\Black{Type3No.2}}
    
    \Photon(280,-280)(310,-300){2}{4}
    \Line[dash,dashsize=5,arrow,arrowpos=0.5,arrowlength=3,arrowwidth=1.1,arrowinset=0.2](310,-260)(280,-280)
    \Line[arrow,arrowpos=0.5,arrowlength=3,arrowwidth=1.1,arrowinset=0.2](310,-260)(310,-300)
    \Line[arrow,arrowpos=0.5,arrowlength=3,arrowwidth=1.1,arrowinset=0.2](340,-250)(310,-260)
    \Line[arrow,arrowpos=0.5,arrowlength=3,arrowwidth=1.1,arrowinset=0.2](310,-300)(340,-310)
    \Line[dash,dashsize=5](280,-280)(250,-280)
    \Text(320,-251)[lb]{\Black{$\tau$}}
    \Text(288,-264)[lb]{\Black{$\phi^{0}_{i}$}}
    \Text(288,-307)[lb]{\Black{$Z$}}
    \Text(315,-283)[lb]{\Black{$\chi^{-}_k$}}
    \Text(320,-318)[lb]{\Black{$\mu$}}
    \Text(260,-295)[lb]{\Black{$\phi^{0}$}}
    \Vertex(310,-260){1.5}
    \Vertex(280,-280){1.5}
    \Vertex(310,-300){1.5}  
    \Text(295,-330)[]{\Black{Type3No.3}}   
         
    \Photon(395,-280)(425,-300){2}{4}
    \Line[dash,dashsize=5,arrow,arrowpos=0.5,arrowlength=3,arrowwidth=1.1,arrowinset=0.2](425,-260)(395,-280)
   \Line[arrow,arrowpos=0.5,arrowlength=3,arrowwidth=1.1,arrowinset=0.2](425,-260)(425,-300)
    \Line[arrow,arrowpos=0.5,arrowlength=3,arrowwidth=1.1,arrowinset=0.2](455,-250)(425,-260)
    \Line[arrow,arrowpos=0.5,arrowlength=3,arrowwidth=1.1,arrowinset=0.2](425,-300)(455,-310)
    \Line[dash,dashsize=5](395,-280)(365,-280)
    \Text(435,-251)[lb]{\Black{$\tau$}}
    \Text(403,-264)[lb]{\Black{$\phi^{-}_{i}$}}
    \Text(403,-308)[lb]{\Black{$W$}}
    \Text(430,-283)[lb]{\Black{$\chi^0_k$}}
    \Text(435,-318)[lb]{\Black{$\mu$}}
    \Text(375,-295)[lb]{\Black{$\phi^{0}$}}
    \Vertex(425,-260){1.5}
    \Vertex(395,-280){1.5}
    \Vertex(425,-300){1.5}
    \Text(410,-330)[]{\Black{Type3No.4}}
    
    \Line[arrow,arrowpos=0.5,arrowlength=3,arrowwidth=1.1,arrowinset=0.2](80,-380)(50,-400)
    \Line[arrow,arrowpos=0.5,arrowlength=3,arrowwidth=1.1,arrowinset=0.2](50,-400)(80,-420)
    \Photon(80,-380)(80,-420){2}{4}
    \Line[arrow,arrowpos=0.5,arrowlength=3,arrowwidth=1.1,arrowinset=0.2](110,-370)(80,-380)
    \Line[arrow,arrowpos=0.5,arrowlength=3,arrowwidth=1.1,arrowinset=0.2](80,-420)(110,-430)
    \Line[dash,dashsize=5](50,-400)(20,-400)
    \Text(90,-371)[lb]{\Black{$\tau$}}
    \Text(60,-385)[lb]{\Black{$\chi^-_i$}}
    \Text(59,-430)[lb]{\Black{$\chi^-_j$}}
    \Text(85,-404)[lb]{\Black{$Z$}}
    \Text(90,-439)[lb]{\Black{$\mu$}}
    \Text(30,-416)[lb]{\Black{$\phi^{0}$}}
    \Vertex(80,-380){1.5}
    \Vertex(50,-400){1.5}
    \Vertex(80,-420){1.5}  
    \Text(65,-450)[]{\Black{Type4No.1}} 
        
    \Line[arrow,arrowpos=0.5,arrowlength=3,arrowwidth=1.1,arrowinset=0.2](195,-380)(165,-400)
    \Line[arrow,arrowpos=0.5,arrowlength=3,arrowwidth=1.1,arrowinset=0.2](165,-400)(195,-420)
    \Photon(195,-380)(195,-420){2}{4}
    \Line[arrow,arrowpos=0.5,arrowlength=3,arrowwidth=1.1,arrowinset=0.2](225,-370)(195,-380)
    \Line[arrow,arrowpos=0.5,arrowlength=3,arrowwidth=1.1,arrowinset=0.2](195,-420)(225,-430)
    \Line[dash,dashsize=5](165,-400)(135,-400)
    \Text(205,-371)[lb]{\Black{$\tau$}}
    \Text(175,-384)[lb]{\Black{$\chi^0_i$}}
    \Text(174,-430)[lb]{\Black{$\chi^0_j$}}
    \Text(200,-403)[lb]{\Black{$W$}}
    \Text(205,-439)[lb]{\Black{$\mu$}}
    \Text(145,-416)[lb]{\Black{$\phi^{0}$}}
    \Vertex(195,-380){1.5}
    \Vertex(165,-400){1.5}
    \Vertex(195,-420){1.5}
    \Text(180,-450)[]{\Black{Type4No.2}}
     
    \Line[arrow,arrowpos=0.5,arrowlength=3,arrowwidth=1.1,arrowinset=0.2](310,-380)(280,-400)
    \Line[arrow,arrowpos=0.5,arrowlength=3,arrowwidth=1.1,arrowinset=0.2](280,-400)(310,-420)
    \Photon(310,-380)(310,-420){2}{4}
    \Line[arrow,arrowpos=0.5,arrowlength=3,arrowwidth=1.1,arrowinset=0.2](340,-370)(310,-380)
    \Line[arrow,arrowpos=0.5,arrowlength=3,arrowwidth=1.1,arrowinset=0.2](310,-420)(340,-430)
    \Line[dash,dashsize=5](280,-400)(250,-400)
    \Text(320,-371)[lb]{\Black{$\tau$}}
    \Text(290,-385)[lb]{\Black{$\chi^-_i$}}
    \Text(288,-429)[lb]{\Black{$\chi^-_j$}}
    \Text(315,-404)[lb]{\Black{$\gamma$}}
    \Text(320,-439)[lb]{\Black{$\mu$}}
    \Text(260,-416)[lb]{\Black{$\phi^{0}$}}
    \Vertex(310,-380){1.5}
    \Vertex(280,-400){1.5}
    \Vertex(310,-420){1.5}  
    \Text(295,-450)[]{\Black{Type4No.3}} 
    
    \Photon(80,-500)(50,-520){2}{4}
    \Photon(50,-520)(80,-540){2}{4}
    \Line[arrow,arrowpos=0.5,arrowlength=3,arrowwidth=1.1,arrowinset=0.2](80,-500)(80,-540)
    \Line[arrow,arrowpos=0.5,arrowlength=3,arrowwidth=1.1,arrowinset=0.2](110,-490)(80,-500)
    \Line[arrow,arrowpos=0.5,arrowlength=3,arrowwidth=1.1,arrowinset=0.2](80,-540)(110,-550)
    \Line[dash,dashsize=5](50,-520)(20,-520)
    \Text(90,-491)[lb]{\Black{$\tau$}}
    \Text(58,-504)[lb]{\Black{$W$}}
    \Text(57,-549)[lb]{\Black{$W$}}
    \Text(85,-523)[lb]{\Black{$\chi^0_k$}}
    \Text(90,-559)[lb]{\Black{$\mu$}}
    \Text(30,-536)[lb]{\Black{$\phi^{0}$}}
    \Vertex(80,-500){1.5}
    \Vertex(50,-520){1.5}
    \Vertex(80,-540){1.5}
    \Text(65,-570)[]{\Black{Type5No.1}}
              
  \end{picture}

\end{center}

\newpage

\begin{center}

  \begin{picture}(450,85)(20,-70)
    \SetWidth{0.5}
    \SetColor{Black}
 
    \Line[arrow,arrowpos=0.5,arrowlength=3,arrowwidth=1.1,arrowinset=0.2](110,-10)(60,-40)
    \Line[arrow,arrowpos=0.5,arrowlength=3,arrowwidth=1.1,arrowinset=0.2](60,-40)(80,-52)
    \Arc[arrow,arrowpos=0.5,arrowlength=3,arrowwidth=1.1,arrowinset=0.2](89,-53.389)(9.5,171.227,308.302)
   \Arc[dash,dashsize=5,arrow,arrowpos=0.5,arrowlength=3,arrowwidth=1.1,arrowinset=0.2](85.3,-59.611)(9.5,-8.773,125)
    \Line[arrow,arrowpos=0.5,arrowlength=3,arrowwidth=1.1,arrowinset=0.2](95,-61)(110,-70)
    \Line[dash,dashsize=5](60,-40)(20,-40)
    \Vertex(60,-40){1.5}
    \Vertex(95,-61){1.5}
    \Vertex(80,-52){1.5}   
    \Text(80,-20)[lb]{\Black{$\tau$}}
    \Text(104,-62)[lb]{\Black{$\mu$}}
    \Text(35,-54)[lb]{\Black{$\phi^{0}$}}
    \Text(91,-50)[lb]{\Black{$\tilde{u}_{j}$}}
    \Text(60,-61)[lb]{\Black{$\chi^{-}_{i}$}}
    \Text(77,-76)[lb]{\Black{$d_{k}$}}
    \Text(65,-90)[]{\Black{Type6No.1}}
    
    \Line[arrow,arrowpos=0.5,arrowlength=3,arrowwidth=1.1,arrowinset=0.2](225,-10)(175,-40)
    \Line[arrow,arrowpos=0.5,arrowlength=3,arrowwidth=1.1,arrowinset=0.2](175,-40)(195,-52)
    \Arc[arrow,arrowpos=0.5,arrowlength=3,arrowwidth=1.1,arrowinset=0.2,clock](204,-53.389)(9.5,308.302,171.227)
   \Arc[dash,dashsize=5,arrow,arrowpos=0.5,arrowlength=3,arrowwidth=1.1,arrowinset=0.2,clock](200.3,-59.611)(9.5,125,-8.773)
    \Line[arrow,arrowpos=0.5,arrowlength=3,arrowwidth=1.1,arrowinset=0.2](210,-61)(225,-70)
   \Line[dash,dashsize=5](175,-40)(135,-40)
    \Vertex(175,-40){1.5}
    \Vertex(210,-61){1.5}
    \Vertex(195,-52){1.5}   
    \Text(195,-20)[lb]{\Black{$\tau$}}
    \Text(219,-62)[lb]{\Black{$\mu$}}
    \Text(150,-54)[lb]{\Black{$\phi^{0}$}}
    \Text(206,-50)[lb]{\Black{$\tilde{d}_{j}$}}
    \Text(175,-61)[lb]{\Black{$\chi^{-}_{i}$}}
    \Text(192,-74)[lb]{\Black{$u_{k}$}}
    \Text(180,-90)[]{\Black{Type6No.2}}
    
    \Line[arrow,arrowpos=0.5,arrowlength=3,arrowwidth=1.1,arrowinset=0.2](340,-10)(290,-40)
    \Line[arrow,arrowpos=0.5,arrowlength=3,arrowwidth=1.1,arrowinset=0.2](290,-40)(310,-52)
    \Arc[arrow,arrowpos=0.5,arrowlength=3,arrowwidth=1.1,arrowinset=0.2](319,-53.389)(9.5,171.227,308.302)
   \Arc[dash,dashsize=5,arrow,arrowpos=0.5,arrowlength=3,arrowwidth=1.1,arrowinset=0.2,clock](315.3,-59.611)(9.5,125,-8.773)
    \Line[arrow,arrowpos=0.5,arrowlength=3,arrowwidth=1.1,arrowinset=0.2](325,-61)(340,-70)
    \Line[dash,dashsize=5](290,-40)(250,-40)
    \Vertex(290,-40){1.5}
    \Vertex(325,-61){1.5}
    \Vertex(310,-52){1.5}   
    \Text(310,-20)[lb]{\Black{$\tau$}}
    \Text(334,-62)[lb]{\Black{$\mu$}}
    \Text(265,-54)[lb]{\Black{$\phi^{0}$}}
    \Text(321,-50)[lb]{\Black{$\phi^{0}_{j}$}}
    \Text(290,-61)[lb]{\Black{$\chi^{-}_{i}$}}
    \Text(305,-77)[lb]{\Black{$\chi^{-}_{k}$}}
    \Text(295,-90)[]{\Black{Type6No.3}}
    
    \Line[arrow,arrowpos=0.5,arrowlength=3,arrowwidth=1.1,arrowinset=0.2](455,-10)(405,-40)
    \Line[arrow,arrowpos=0.5,arrowlength=3,arrowwidth=1.1,arrowinset=0.2](405,-40)(425,-52)
    \Arc[arrow,arrowpos=0.5,arrowlength=3,arrowwidth=1.1,arrowinset=0.2](434,-53.389)(9.5,171.227,308.302)
   \Arc[dash,dashsize=5,arrow,arrowpos=0.5,arrowlength=3,arrowwidth=1.1,arrowinset=0.2,clock](430.3,-59.611)(9.5,125,-8.773)
    \Line[arrow,arrowpos=0.5,arrowlength=3,arrowwidth=1.1,arrowinset=0.2](440,-61)(455,-70)
    \Line[dash,dashsize=5](405,-40)(365,-40)
    \Vertex(405,-40){1.5}
    \Vertex(440,-61){1.5}
    \Vertex(425,-52){1.5}   
    \Text(425,-20)[lb]{\Black{$\tau$}}
    \Text(449,-62)[lb]{\Black{$\mu$}}
    \Text(380,-54)[lb]{\Black{$\phi^{0}$}}
    \Text(436,-50)[lb]{\Black{$\phi^{-}_{j}$}}
    \Text(405,-61)[lb]{\Black{$\chi^{-}_{i}$}}
    \Text(420,-78)[lb]{\Black{$\chi^{0}_{k}$}}
    \Text(410,-90)[]{\Black{Type6No.4}}

    \Line[arrow,arrowpos=0.5,arrowlength=3,arrowwidth=1.1,arrowinset=0.2](110,-130)(95,-139)
    \Line[arrow,arrowpos=0.5,arrowlength=3,arrowwidth=1.1,arrowinset=0.2](80,-148)(60,-160)
    \Arc[arrow,arrowpos=0.5,arrowlength=3,arrowwidth=1.1,arrowinset=0.2](89,-146.389)(9.5,51.698,188.773)  
   \Arc[dash,dashsize=5,arrow,arrowpos=0.5,arrowlength=3,arrowwidth=1.1,arrowinset=0.2](85.45,-140.511)(9.5,-125,8.773)
    \Line[arrow,arrowpos=0.5,arrowlength=3,arrowwidth=1.1,arrowinset=0.2](60,-160)(110,-190)
    \Line[dash,dashsize=5](60,-160)(20,-160)
    \Vertex(60,-160){1.5}
    \Vertex(95,-139){1.5}
    \Vertex(80,-148){1.5}   
    \Text(99,-131)[lb]{\Black{$\tau$}}
    \Text(79,-187)[lb]{\Black{$\mu$}}
    \Text(35,-175)[lb]{\Black{$\phi^{0}$}}
    \Text(91,-164)[lb]{\Black{$\tilde{u}_{j}$}}
    \Text(63,-150)[lb]{\Black{$\chi^{-}_{i}$}}
    \Text(76,-135)[lb]{\Black{$d_{k}$}}
    \Text(65,-210)[]{\Black{Type7No.1}}
    
    \Line[arrow,arrowpos=0.5,arrowlength=3,arrowwidth=1.1,arrowinset=0.2](225,-130)(210,-139)
    \Line[arrow,arrowpos=0.5,arrowlength=3,arrowwidth=1.1,arrowinset=0.2](195,-148)(175,-160)
    \Arc[arrow,arrowpos=0.5,arrowlength=3,arrowwidth=1.1,arrowinset=0.2,clock](204,-146.389)(9.5,188.773,51.698)  
   \Arc[dash,dashsize=5,arrow,arrowpos=0.5,arrowlength=3,arrowwidth=1.1,arrowinset=0.2,clock](200.45,-140.511)(9.5,8.773,-125)
    \Line[arrow,arrowpos=0.5,arrowlength=3,arrowwidth=1.1,arrowinset=0.2](175,-160)(225,-190)
    \Line[dash,dashsize=5](175,-160)(135,-160)
    \Vertex(175,-160){1.5}
    \Vertex(210,-139){1.5}
    \Vertex(195,-148){1.5}   
    \Text(214,-131)[lb]{\Black{$\tau$}}
    \Text(194,-187)[lb]{\Black{$\mu$}}
    \Text(150,-175)[lb]{\Black{$\phi^{0}$}}
    \Text(206,-167)[lb]{\Black{$\tilde{d}_{j}$}}
    \Text(178,-150)[lb]{\Black{$\chi^{-}_{i}$}}
    \Text(191,-135)[lb]{\Black{$u_{k}$}}
    \Text(180,-210)[]{\Black{Type7No.2}}
    
    \Line[arrow,arrowpos=0.5,arrowlength=3,arrowwidth=1.1,arrowinset=0.2](340,-130)(325,-139)
    \Line[arrow,arrowpos=0.5,arrowlength=3,arrowwidth=1.1,arrowinset=0.2](310,-148)(290,-160)
    \Arc[arrow,arrowpos=0.5,arrowlength=3,arrowwidth=1.1,arrowinset=0.2](319,-146.389)(9.5,51.698,188.773)  
   \Arc[dash,dashsize=5,arrow,arrowpos=0.5,arrowlength=3,arrowwidth=1.1,arrowinset=0.2,clock](315.45,-140.511)(9.5,8.773,-125)
    \Line[arrow,arrowpos=0.5,arrowlength=3,arrowwidth=1.1,arrowinset=0.2](290,-160)(340,-190)
    \Line[dash,dashsize=5](290,-160)(250,-160)
    \Vertex(290,-160){1.5}
    \Vertex(325,-139){1.5}
    \Vertex(310,-148){1.5}   
    \Text(329,-131)[lb]{\Black{$\tau$}}
    \Text(309,-187)[lb]{\Black{$\mu$}}
    \Text(265,-175)[lb]{\Black{$\phi^{0}$}}
    \Text(322,-164)[lb]{\Black{$\phi^{0}_{j}$}}
    \Text(293,-150)[lb]{\Black{$\chi^{-}_{i}$}}
    \Text(306,-135)[lb]{\Black{$\chi^{-}_{k}$}}
    \Text(295,-210)[]{\Black{Type7No.3}}
    
    \Line[arrow,arrowpos=0.5,arrowlength=3,arrowwidth=1.1,arrowinset=0.2](455,-130)(440,-139)
    \Line[arrow,arrowpos=0.5,arrowlength=3,arrowwidth=1.1,arrowinset=0.2](425,-148)(405,-160)
    \Arc[arrow,arrowpos=0.5,arrowlength=3,arrowwidth=1.1,arrowinset=0.2](434,-146.389)(9.5,51.698,188.773)  
   \Arc[dash,dashsize=5,arrow,arrowpos=0.5,arrowlength=3,arrowwidth=1.1,arrowinset=0.2,clock](430.45,-140.511)(9.5,8.773,-125)
    \Line[arrow,arrowpos=0.5,arrowlength=3,arrowwidth=1.1,arrowinset=0.2](405,-160)(455,-190)
    \Line[dash,dashsize=5](405,-160)(365,-160)
    \Vertex(405,-160){1.5}
    \Vertex(440,-139){1.5}
    \Vertex(425,-148){1.5}   
    \Text(444,-131)[lb]{\Black{$\tau$}}
    \Text(424,-187)[lb]{\Black{$\mu$}}
    \Text(380,-175)[lb]{\Black{$\phi^{0}$}}
    \Text(437,-164)[lb]{\Black{$\phi^{-}_{j}$}}
    \Text(408,-150)[lb]{\Black{$\chi^{-}_{i}$}}
    \Text(421,-135)[lb]{\Black{$\chi^{0}_{k}$}}
    \Text(410,-210)[]{\Black{Type7No.4}}

    \Line[arrow,arrowpos=0.5,arrowlength=3,arrowwidth=1.1,arrowinset=0.2](110,-250)(60,-280)
    \Line[arrow,arrowpos=0.5,arrowlength=3,arrowwidth=1.1,arrowinset=0.2](60,-280)(80,-292)
    \Arc[arrow,arrowpos=0.5,arrowlength=3,arrowwidth=1.1,arrowinset=0.2](89,-293.389)(9.5,171.227,308.302)
   \PhotonArc[clock](85.3,-299.611)(9.5,125,-8.773){1.6}{4}
    \Line[arrow,arrowpos=0.5,arrowlength=3,arrowwidth=1.1,arrowinset=0.2](95,-301)(110,-310)
    \Line[dash,dashsize=5](60,-280)(20,-280)
    \Vertex(60,-280){1.5}
    \Vertex(95,-301){1.5}
    \Vertex(80,-292){1.5}   
    \Text(80,-260)[lb]{\Black{$\tau$}}
    \Text(104,-302)[lb]{\Black{$\mu$}}
    \Text(35,-296)[lb]{\Black{$\phi^{0}$}}
    \Text(92,-289)[lb]{\Black{$Z$}}
    \Text(60,-301)[lb]{\Black{$\chi^-_i$}}
    \Text(77,-317)[lb]{\Black{$\chi^-_k$}}
    \Text(65,-330)[]{\Black{Type8No.1}}
    
    \Line[arrow,arrowpos=0.5,arrowlength=3,arrowwidth=1.1,arrowinset=0.2](225,-250)(175,-280)
    \Line[arrow,arrowpos=0.5,arrowlength=3,arrowwidth=1.1,arrowinset=0.2](175,-280)(195,-292)
    \Arc[arrow,arrowpos=0.5,arrowlength=3,arrowwidth=1.1,arrowinset=0.2](204,-293.389)(9.5,171.227,308.302)
   \PhotonArc[clock](200.3,-299.611)(9.5,125,-8.773){1.6}{4}
    \Line[arrow,arrowpos=0.5,arrowlength=3,arrowwidth=1.1,arrowinset=0.2](210,-301)(225,-310)
    \Line[dash,dashsize=5](175,-280)(135,-280)
    \Vertex(175,-280){1.5}
    \Vertex(210,-301){1.5}
    \Vertex(195,-292){1.5}   
    \Text(195,-260)[lb]{\Black{$\tau$}}
    \Text(219,-302)[lb]{\Black{$\mu$}}
    \Text(150,-296)[lb]{\Black{$\phi^{0}$}}
    \Text(207,-289)[lb]{\Black{$W$}}
    \Text(175,-301)[lb]{\Black{$\chi^{-}_i$}}
    \Text(192,-319)[lb]{\Black{$\chi^0_k$}}
    \Text(180,-330)[]{\Black{Type8No.2}}

    \Line[arrow,arrowpos=0.5,arrowlength=3,arrowwidth=1.1,arrowinset=0.2](340,-250)(290,-280)
    \Line[arrow,arrowpos=0.5,arrowlength=3,arrowwidth=1.1,arrowinset=0.2](290,-280)(310,-292)
    \Arc[arrow,arrowpos=0.5,arrowlength=3,arrowwidth=1.1,arrowinset=0.2](319,-293.389)(9.5,171.227,308.302)
   \PhotonArc[clock](315.3,-299.611)(9.5,125,-8.773){1.6}{4}
    \Line[arrow,arrowpos=0.5,arrowlength=3,arrowwidth=1.1,arrowinset=0.2](325,-301)(340,-310)
    \Line[dash,dashsize=5](290,-280)(250,-280)
    \Vertex(290,-280){1.5}
    \Vertex(325,-301){1.5}
    \Vertex(310,-292){1.5}   
    \Text(310,-260)[lb]{\Black{$\tau$}}
    \Text(334,-302)[lb]{\Black{$\mu$}}
    \Text(265,-296)[lb]{\Black{$\phi^{0}$}}
    \Text(322,-289)[lb]{\Black{$\gamma$}}
    \Text(290,-301)[lb]{\Black{$\chi^-_i$}}
    \Text(304,-316)[lb]{\Black{$\chi^-_k$}}
    \Text(295,-330)[]{\Black{Type8No.3}}
    
    \Line[arrow,arrowpos=0.5,arrowlength=3,arrowwidth=1.1,arrowinset=0.2](110,-370)(95,-379)
    \Line[arrow,arrowpos=0.5,arrowlength=3,arrowwidth=1.1,arrowinset=0.2](80,-388)(60,-400)
    \Arc[arrow,arrowpos=0.5,arrowlength=3,arrowwidth=1.1,arrowinset=0.2](89,-386.389)(9.5,51.698,188.773)  
   \PhotonArc[clock](85.45,-380.511)(9.5,8.773,-125){1.6}{4}
    \Line[arrow,arrowpos=0.5,arrowlength=3,arrowwidth=1.1,arrowinset=0.2](60,-400)(110,-430)
    \Line[dash,dashsize=5](60,-400)(20,-400)
    \Vertex(60,-400){1.5}
    \Vertex(95,-379){1.5}
    \Vertex(80,-388){1.5}   
    \Text(99,-371)[lb]{\Black{$\tau$}}
    \Text(79,-428)[lb]{\Black{$\mu$}}
    \Text(35,-416)[lb]{\Black{$\phi^{0}$}}
    \Text(91,-402)[lb]{\Black{$Z$}}
    \Text(61,-391)[lb]{\Black{$\chi^-_i$}}
    \Text(76,-376)[lb]{\Black{$\chi^-_k$}}
    \Text(65,-450)[]{\Black{Type9No.1}}
    
    \Line[arrow,arrowpos=0.5,arrowlength=3,arrowwidth=1.1,arrowinset=0.2](225,-370)(210,-379)
    \Line[arrow,arrowpos=0.5,arrowlength=3,arrowwidth=1.1,arrowinset=0.2](195,-388)(175,-400)
    \Arc[arrow,arrowpos=0.5,arrowlength=3,arrowwidth=1.1,arrowinset=0.2](204,-386.389)(9.5,51.698,188.773)  
   \PhotonArc[clock](200.45,-380.511)(9.5,8.773,-125){1.6}{4}
    \Line[arrow,arrowpos=0.5,arrowlength=3,arrowwidth=1.1,arrowinset=0.2](175,-400)(225,-430)
    \Line[dash,dashsize=5](175,-400)(135,-400)
    \Vertex(175,-400){1.5}
    \Vertex(210,-379){1.5}
    \Vertex(195,-388){1.5}   
    \Text(214,-371)[lb]{\Black{$\tau$}}
    \Text(194,-428)[lb]{\Black{$\mu$}}
    \Text(150,-416)[lb]{\Black{$\phi^{0}$}}
    \Text(206,-403)[lb]{\Black{$W$}}
    \Text(176,-391)[lb]{\Black{$\chi^{-}_i$}}
    \Text(191,-375)[lb]{\Black{$\chi^0_k$}}
    \Text(180,-450)[]{\Black{Type9No.2}}

    \Line[arrow,arrowpos=0.5,arrowlength=3,arrowwidth=1.1,arrowinset=0.2](340,-370)(325,-379)
    \Line[arrow,arrowpos=0.5,arrowlength=3,arrowwidth=1.1,arrowinset=0.2](310,-388)(290,-400)
    \Arc[arrow,arrowpos=0.5,arrowlength=3,arrowwidth=1.1,arrowinset=0.2](319,-386.389)(9.5,51.698,188.773)  
   \PhotonArc[clock](315.45,-380.511)(9.5,8.773,-125){1.6}{4}
    \Line[arrow,arrowpos=0.5,arrowlength=3,arrowwidth=1.1,arrowinset=0.2](290,-400)(340,-430)
    \Line[dash,dashsize=5](290,-400)(250,-400)
    \Vertex(290,-400){1.5}
    \Vertex(325,-379){1.5}
    \Vertex(310,-388){1.5}   
    \Text(329,-371)[lb]{\Black{$\tau$}}
    \Text(309,-428)[lb]{\Black{$\mu$}}
    \Text(265,-416)[lb]{\Black{$\phi^{0}$}}
    \Text(322,-402)[lb]{\Black{$\gamma$}}
    \Text(295,-391)[lb]{\Black{$\chi^-_i$}}
    \Text(308,-376)[lb]{\Black{$\chi^-_k$}}
    \Text(295,-450)[]{\Black{Type9No.3}}
    
  \end{picture}

\end{center}
\newpage
\section{Effective Couplings in MSSM without \mbox{\boldmath $R$} parity}
We list all relevant effective mass eigenstate couplings for our analysis here.
Indices run from 1 to 10 for neutral scalars (sleptons), 1 to 8 for charged scalars (sleptons), 1 to 6 for squarks, 1 to 7 for neutral fermions (neutralinos) and 1 to 5 for charged fermions (charginos) while all dummy indices run from 1 to 3. Moreover,
\[
y_{u_i}=\frac{g_{\scriptscriptstyle 2}m_{u_i}}{\sqrt{2}M_{W}\sin\!\beta}\,,\hspace{5mm}
y_{d_i}=\frac{g_{\scriptscriptstyle 2}m_{d_i}}{\sqrt{2}M_{W}\cos\!\beta}\,,\hspace{5mm} \text{and}\hspace{5mm}
y_{e_i}=\frac{g_{\scriptscriptstyle 2}m_i}{\sqrt{2}M_{W}\cos\!\beta} 
\]
are the diagonal quark and charged lepton Yukawa couplings, where $m_i$'s ($\approx m_{e_i}$ under the small-$\mu_i$ scenario) are mass parameters in the charged fermion mass matrix \cite{otto}.

\bigskip\medskip
\noindent{\bf Neutral Scalar (Sneutrino)-\mbox{\boldmath $W^+$}-\mbox{\boldmath $W^-$} Vertices}
\[{\cal L}=g^{\scriptscriptstyle W}_m W^+ W^-\phi^{0}_{m}\]
where
\begin{equation}g^{\scriptscriptstyle W}_m=
g_{\scriptscriptstyle 2}M_W\left(\sin\!\beta\,{\cal D}^{s}_{1m}+\cos\!\beta\,{\cal D}^{s}_{2m}\right) \,. 
\end{equation}

\bigskip\medskip
\noindent{\bf Neutral Scalar-Chagred Scalar (Slepton)-\mbox{\boldmath $W$} Vertices}
\[
{\cal L}={\cal G}^{\scriptscriptstyle W}_{mn}\left[p(\phi^{-}_{n})-p(\phi^{0}_{m})\right]_{\mu}
  W^+\phi^-_n\phi^0_m  +\text{h.c.}
\]
where
\begin{equation}
{\cal G}^{\scriptscriptstyle W}_{mn}=
\frac{1}{2}g_{\scriptscriptstyle 2}\left[\left({\cal D}^{s}_{1m}-i{\cal D}^{s}_{6m}\right){\cal D}^{l}_{1n}
  -\left({\cal D}^{s}_{2m}-i{\cal D}^{s}_{7m}\right){\cal D}^{l}_{2n}
  -\left({\cal D}^{s}_{\left(p+2\right)m}-i{\cal D}^{s}_{\left(p+7\right)m}\right){\cal D}^{l}_{\left(p+2\right)n}\right]\,.
\end{equation}

\bigskip\medskip
\noindent{\bf Neutral Scalar-Neutral Scalar-\mbox{\boldmath $Z$} Vertices}
\[{\cal L}={\cal G}^{\scriptscriptstyle Z}_{ij}
\left[p(\phi^{0}_{j})-p(\phi^{0}_{i})\right]_{\mu}Z\phi^{0}_{i}\phi^{0}_{j}
\]
where
\begin{equation}{\cal G}^{\scriptscriptstyle Z}_{ij}=
\frac{i}{2}g_{\scriptscriptstyle Z}\!
\left({\cal D}^{s}_{6i}{\cal D}^{s}_{1j}-{\cal D}^{s}_{1i}{\cal D}^{s}_{6j}  
+{\cal D}^{s}_{7i}{\cal D}^{s}_{2j}-{\cal D}^{s}_{2i}{\cal D}^{s}_{7j}
+{\cal D}^{s}_{\left(q+7\right)i}{\cal D}^{s}_{\left(q+2\right)j}  
-{\cal D}^{s}_{\left(q+2\right)i}{\cal D}^{s}_{\left(q+7\right)j}\right)\,.
\end{equation}

\bigskip\medskip
\noindent{\bf Neutral Lepton (Neutralino)-Charged Lepton (Chargino)-\mbox{\boldmath $W$} Vertices}
\[
{\cal L}=
 g_{\scriptscriptstyle 2}\overline{\Psi}(\chi^0_m)\Phi^\dagger(W^-)
 \left[{\cal N}^{wL}_{mi}\gamma_\mu\frac{1-\gamma_5}{2}
      +{\cal N}^{wR}_{mi}\gamma_\mu\frac{1+\gamma_5}{2} \right]\Psi(\chi^-_i)+\text{h.c.}  \]
where   
\begin{align}   
{\cal N}^{wL}_{mi}&=
 -\mbox{\boldmath $X$}^*_{\!2m}  \mbox{\boldmath $U$}_{\!1i}
 -\frac{1}{\sqrt{2}}\mbox{\boldmath $X$}^*_{\!4m} \mbox{\boldmath $U$}_{\!2i}
 -\frac{1}{\sqrt{2}}\mbox{\boldmath $X$}^*_{\!(q+4)m} \mbox{\boldmath $U$}_{\!\left(q+2\right)i}
  \nonumber\\
{\cal N}^{wR}_{mi}&=   
  \mbox{\boldmath $V$}_{\!1i}   \mbox{\boldmath $X$}_{\!2m}
 +\frac{1}{\sqrt{2}}\mbox{\boldmath $V$}_{\!2i} \mbox{\boldmath $X$}^*_{\!3m}\,.  
\end{align}

\bigskip\medskip
\noindent{\bf Charged Lepton-Charged Lepton-\mbox{\boldmath $Z$} Vertices}
\[
{\cal L}=g_{\scriptscriptstyle Z}\overline{\Psi}(\chi^-_i)\left[ 
 {\cal C}^{zL}_{ij}\gamma_\mu\frac{1-\gamma_5}{2} 
+{\cal C}^{zR}_{ij}\gamma_\mu\frac{1+\gamma_5}{2}\right] \Phi(Z) \Psi(\chi^-_j)  
\]
where
\begin{align}
{\cal C}^{zL}_{ij}&=
-\left(-1+\sin^2\!\theta_{\scriptscriptstyle W}\right)
  \mbox{\boldmath $U$}^*_{1i}  \mbox{\boldmath $U$}_{\!1j} 
-\left(-\frac{1}{2}+\sin^{2}\!\theta_{\scriptscriptstyle W}\right)
  \mbox{\boldmath $U$}^*_{2i} \mbox{\boldmath $U$}_{\!2j} 
-\left(-\frac{1}{2}+\sin^{2}\!\theta_{\scriptscriptstyle W}\right)
  \mbox{\boldmath $U$}^*_{\left(q+2\right)i} \mbox{\boldmath $U$}_{\!\left(q+2\right)j} \nonumber\\
{\cal C}^{zR}_{ij}&=
\left(1-\sin^2\!\theta_{\scriptscriptstyle W}\right)
  \mbox{\boldmath $V$}_{\!1j} \mbox{\boldmath $V$}^*_{1i} 
+\left(\frac{1}{2}-\sin^{2}\!\theta_{\scriptscriptstyle W}\right)
  \mbox{\boldmath $V$}_{\!2j}  \mbox{\boldmath $V$}^*_{2i}  
-\sin^{2}\!\theta_{\scriptscriptstyle W}\mbox{\boldmath $V$}_{\!\left(q+2\right)j}  
  \mbox{\boldmath $V$}^*_{\left(q+2\right)i} \;. 
\end{align}

\bigskip\medskip
\noindent{\bf Charged Lepton-Charged Lepton-\mbox{\boldmath $\gamma$} Vertices}
\begin{align}
{\cal L}=e\overline{\Psi}(\chi^-_i) \left[
 {\cal C}^{\gamma L}_{ij}\gamma_\mu\frac{1-\gamma_5}{2}  
+{\cal C}^{\gamma R}_{ij}\gamma_\mu\frac{1+\gamma_5}{2} \right]\Phi(A^\mu)\Psi(\chi^-_j) \nonumber  
\end{align}
where	
\begin{align}
{\cal C}^{\gamma L}_{ij}&=
  \mbox{\boldmath $U$}^*_{1i}   \mbox{\boldmath $U$}_{\!1j}
 +\mbox{\boldmath $U$}^*_{2i} \mbox{\boldmath $U$}_{\!2j}
 +\mbox{\boldmath $U$}^*_{\left(q+2\right)i} \mbox{\boldmath $U$}_{\!\left(q+2\right)j} \nonumber\\
{\cal C}^{\gamma R}_{ij}&=
  \mbox{\boldmath $V$}_{\!1j}  \mbox{\boldmath $V$}^*_{1i} 
 +\mbox{\boldmath $V$}_{\!2j}  \mbox{\boldmath $V$}^*_{2i} 
 +\mbox{\boldmath $V$}_{\!\left(q+2\right)j}  \mbox{\boldmath $V$}^*_{\left(q+2\right)i} \;.
\end{align}	

\bigskip\medskip
\noindent{\bf Charged Lepton-Down Quark-Up Squark Vertices}
\[
{\cal L}^{\!\chi^-}
= g_{\scriptscriptstyle 2}
\overline{\Psi}(\chi_n^-) \Phi^{\dagger}({\tilde{u}_{m}})
\left[ {\cal C}^{dL}_{nmi}\frac{1-\gamma_5}{2} +{\cal C}^{dR}_{nmi}\frac{1+\gamma_5}{2}   \right]
{\Psi}(d_i) +\text{h.c.} \nonumber
\]
where
\begin{align}
{\cal C}^{dL}_{nmi}&=
  -\mbox{\boldmath $V$}^*_{\!1n} {\cal D}^{u*}_{im} 
  +\frac{y_{u_j}}{g_{\scriptscriptstyle 2}}V_{\scriptscriptstyle C\!K\!M}^{ji}
   \mbox{\boldmath $V$}^*_{\!2n}{\cal D}^{u*}_{(j+3)m} \nonumber\\
{\cal C}^{dR}_{nmi}&= 
   \frac{y_{d_i}}{g_{\scriptscriptstyle 2}}\mbox{\boldmath $U$}^*_{\!2n}{\cal D}^{u*}_{im} 
  +\frac{\lambda'^*_{jhi}}{g_{\scriptscriptstyle 2}} \mbox{\boldmath $U$}^*_{\!(j+2)n} {\cal D}^{u*}_{hm}\,.
\end{align}

\bigskip\medskip
\noindent{\bf Neutral Scalar-Quark-Quark Vertices: Down Sector}
\[
{\cal L}^{\phi^0}=g_{\scriptscriptstyle 2}\overline{\Psi}(d_h)\Phi^\dagger(\phi^0_m)\left[
 \widetilde{\cal N}^{dL}_{hmi}\frac{1-\gamma_5}{2}+\widetilde{\cal N}^{dR}_{hmi}\frac{1+\gamma_5}{2}
   \right]\Psi(d_i) \]
where
\begin{align}
\widetilde{\cal N}^{dL}_{hmi}&=
 -\frac{y_{d_i}}{\sqrt{2}g_{\scriptscriptstyle 2}}\delta_{ih} \left({\cal D}^s_{2m}+i{\cal D}^s_{7m}\right)   
 -\frac{\lambda'_{kih}}{\sqrt{2}g_{\scriptscriptstyle 2}} 
  \left({\cal D}^s_{\left(k+2\right)m}+i{\cal D}^s_{\left(k+7\right)m}\right)\nonumber\\  
\widetilde{\cal N}^{dR}_{hmi}&=
 -\frac{y_{d_i}}{\sqrt{2}g_{\scriptscriptstyle 2}} \delta_{ih}\left({\cal D}^s_{2m}-i{\cal D}^s_{7m}\right) 
 -\frac{\lambda'^*_{khi}}{\sqrt{2}g_{\scriptscriptstyle 2}} 
  \left({\cal D}^s_{\left(k+2\right)m}-i{\cal D}^s_{\left(k+7\right)m}\right) \,. 
\end{align}

\bigskip\medskip
\noindent{\bf Neutral Scalar-Quark-Quark Vertices: Up Sector}
\[
{\cal L}^u=g_{\scriptscriptstyle 2}\overline{\Psi}(u_h)\Phi^\dagger(\phi^0_m)
 \left[\widetilde{\cal N}^{uL}_{hmi}\frac{1-\gamma_5}{2}+\widetilde{\cal N}^{uR}_{hmi}\frac{1+\gamma_5}{2}\right]
 \Psi(u_i)\]
where
\begin{align}
\widetilde{\cal N}^{uL}_{hmi}&=
 -\frac{y_{u_i}}{\sqrt{2}g_{\scriptscriptstyle 2}}\delta_{ih}
  \left({\cal D}^s_{1m}-i{\cal D}^s_{6m}\right) \nonumber\\  
\widetilde{\cal N}^{uR}_{hmi}&=
 -\frac{y_{u_i}}{\sqrt{2}g_{\scriptscriptstyle 2}}\delta_{ih}
  \left({\cal D}^s_{1m}+i{\cal D}^s_{6m}\right)  \,.
\end{align}

\bigskip\medskip
\noindent{\bf Charged Lepton-Up Quark-Down Squark Vertices}
\[
{\cal L}^{\!\chi^{\!+}}
= {g_{\scriptscriptstyle 2}}
 \overline{\Psi}(\chi_n^+) \Phi^{\dagger}(\tilde{d}_m)
 \left[ {\cal C}^{uL}_{nmi}\frac{1-\gamma_5}{2}   +{\cal C}^{uR}_{nmi}\frac{1+\gamma_5}{2}   \right]
 {\Psi}(u_i) +\text{h.c.} \]
where
\begin{align}
{\cal C}^{uL}_{nmi}&=
 -V_{\scriptscriptstyle C\!K\!M}^{ip*} \mbox{\boldmath $U$}_{\!1n}{\cal D}^{d*}_{pm}
 +\frac{y_{d_p}}{g_{\scriptscriptstyle 2}}V_{\scriptscriptstyle C\!K\!M}^{ip*}
  \mbox{\boldmath $U$}_{\!2n}{\cal D}^{d*}_{(p+3)m}
 +\frac{\lambda'_{jhp}}{g_{\scriptscriptstyle 2}} V_{\scriptscriptstyle C\!K\!M}^{ih*} 
  \mbox{\boldmath $U$}_{\!(j+2)n}{\cal D}^{d*}_{(p+3)m} \nonumber\\
{\cal C}^{uR}_{nmi}&=
 \frac{y_{u_i}}{g_{\scriptscriptstyle 2}}V_{\scriptscriptstyle C\!K\!M}^{ip*}
  \mbox{\boldmath $V$}_{\!2n} {\cal D}^{d*}_{pm}\,. 
\end{align}

\bigskip\medskip
\noindent{\bf Neutral Scalar-Charged Lepton-Charged Lepton Vertices}
\[
{\cal L}=g_{\scriptscriptstyle 2}\overline{\Psi}(\chi^-_{\bar{n}})
 \left[{\cal C}^{\chi L}_{\bar{n}mn}\frac{1-\gamma_5}{2}
      +{\cal C}^{\chi R}_{\bar{n}mn}\frac{1+\gamma_5}{2} \right] \Psi(\chi^-_n)\Phi(\phi^0_m) 
\]      
where
\begin{align}
{\cal C}^{\chi L}_{\bar{n}mn}=
&-\frac{1}{\sqrt{2}}\mbox{\boldmath $V$}^*_{2\bar{n}}   
        \mbox{\boldmath $U$}_{\!1n}({\cal D}^s_{1m}+i{\cal D}^s_{6m}) 
 -\frac{1}{\sqrt{2}}\mbox{\boldmath $V$}^*_{1\bar{n}}\mbox{\boldmath $U$}_{\!2n}
        ({\cal D}^s_{2m}-i{\cal D}^s_{7m})   \nonumber\\
&-\frac{1}{\sqrt{2}}\mbox{\boldmath $V$}^*_{1\bar{n}}\mbox{\boldmath $U$}_{\!\left(j+2\right)n} 
        ({\cal D}^s_{\left(j+2\right)m}-i{\cal D}^s_{\left(j+7\right)m})   
 -\frac{y_{e_j}}{\sqrt{2}g_{\scriptscriptstyle 2}}\mbox{\boldmath $V$}^*_{\left(j+2\right)\bar{n}}
        \mbox{\boldmath $U$}_{\!\left(j+2\right)n}({\cal D}^s_{2m}+i{\cal D}^s_{7m}) \nonumber\\
&+\frac{y_{e_j}}{\sqrt{2}g_{\scriptscriptstyle 2}}\mbox{\boldmath $V$}^*_{\left(j+2\right)\bar{n}}
        \mbox{\boldmath $U$}_{\!2n} ({\cal D}^s_{\left(j+2\right)m}+i{\cal D}^s_{\left(j+7\right)m}) 
 +\frac{\lambda_{ijk}}{\sqrt{2}g_{\scriptscriptstyle 2}}\mbox{\boldmath $V$}^*_{\left(k+2\right)\bar{n}}
        \mbox{\boldmath $U$}_{\!\left(i+2\right)n}({\cal D}^s_{\left(j+2\right)m}+i{\cal D}^s_{\left(j+7\right)m}) 
\nonumber\\
{\cal C}^{\chi R}_{\bar{n}mn}=
&-\frac{1}{\sqrt{2}}\mbox{\boldmath $U$}^*_{1\bar{n}} \mbox{\boldmath $V$}_{\!2n}
        ({\cal D}^s_{1m}-i{\cal D}^s_{6m}) 
 -\frac{1}{\sqrt{2}}\mbox{\boldmath $U$}^*_{2\bar{n}}\mbox{\boldmath $V$}_{\!1n}   
        ({\cal D}^s_{2m}+i{\cal D}^s_{7m}) \nonumber\\
&-\frac{1}{\sqrt{2}}\mbox{\boldmath $U$}^*_{\left(j+2\right)\bar{n}} \mbox{\boldmath $V$}_{\!1n}  
        ({\cal D}^s_{\left(j+2\right)m}+i{\cal D}^s_{\left(j+7\right)m})
 -\frac{y_{e_j}}{\sqrt{2}g_{\scriptscriptstyle 2}} \mbox{\boldmath $U$}^*_{\left(j+2\right)\bar{n}}
        \mbox{\boldmath $V$}_{\!\left(j+2\right)n}({\cal D}^s_{2m}-i{\cal D}^s_{7m}) \nonumber\\
&+\frac{y_{e_j}}{\sqrt{2}g_{\scriptscriptstyle 2}}\mbox{\boldmath $U$}^*_{2\bar{n}} 
        \mbox{\boldmath $V$}_{\!\left(j+2\right)n}({\cal D}^s_{\left(j+2\right)m}-i{\cal D}^s_{\left(j+7\right)m})
        \nonumber\\ 
&+\frac{\lambda^*_{ijk}}{\sqrt{2}g_{\scriptscriptstyle 2}} \mbox{\boldmath $U$}^*_{\left(i+2\right)\bar{n}} 
        \mbox{\boldmath $V$}_{\!\left(k+2\right)n}
        ({\cal D}^s_{\left(j+2\right)m}-i{\cal D}^s_{\left(j+7\right)m}) \,.
\end{align}      
  
\bigskip\medskip
\noindent{\bf Neutral Scalar-Neutral Lepton-Neutral Lepton Vertices}
\[
{\cal L}=g_{\scriptscriptstyle 2}\overline{\Psi}(\chi^0_{\bar{n}})\left[
         {\cal N}^{\chi L}_{\bar{n}mn}\frac{1-\gamma_5}{2}
        +{\cal N}^{\chi R}_{\bar{n}mn}\frac{1+\gamma_5}{2}  \right]\Psi(\chi^0_n)\Phi(\phi^0_m) 
\]
where
\begin{align}
{\cal N}^{\chi L}_{\bar{n}mn}&=\hspace*{10pt}
\frac{1}{2}
  \left(-\tan\!\theta_{\scriptscriptstyle W} \mbox{\boldmath $X$}_{\!1\bar{n}}
        + \mbox{\boldmath $X$}_{\!2\bar{n}}   \right)
  \mbox{\boldmath $X$}^*_{\!3n} ({\cal D}^s_{1m}+i{\cal D}^s_{6m})\nonumber\\ 
&\quad+\frac{1}{2}\left(\tan\!\theta_{\scriptscriptstyle W} \mbox{\boldmath $X$}_{\!1\bar{n}}
                 - \mbox{\boldmath $X$}_{\!2\bar{n}} \right)
  \mbox{\boldmath $X$}_{\!4n}({\cal D}^s_{2m}-i{\cal D}^s_{7m}) \nonumber \\  
&\quad+\frac{1}{2}
  \left(\tan\!\theta_{\scriptscriptstyle W}\mbox{\boldmath $X$}_{\!1\bar{n}}-\mbox{\boldmath $X$}_{\!2\bar{n}} \right)
  \mbox{\boldmath $X$}_{\!(k+4)n}({\cal D}^s_{\left(k+2\right)m}-i{\cal D}^s_{\left(k+7\right)m})\nonumber \\
{\cal N}^{\chi R}_{\bar{n}mn}&=\hspace*{10pt}
\frac{1}{2}
  \mbox{\boldmath $X$}_{\!3\bar{n}} \left(-\tan\!\theta_{\scriptscriptstyle W}\mbox{\boldmath $X$}^*_{\!1n}
                                          + \mbox{\boldmath $X$}^*_{\!2n}  \right)
  ({\cal D}^s_{1m}-i{\cal D}^s_{6m})\nonumber\\
&\quad+\frac{1}{2}  \mbox{\boldmath $X$}^*_{\!4\bar{n}}
  \left(\tan\!\theta_{\scriptscriptstyle W}\mbox{\boldmath $X$}^*_{\!1n} -\mbox{\boldmath $X$}^*_{\!2n}  \right)
  ({\cal D}^s_{2m}+i{\cal D}^s_{7m})\nonumber\\
&\quad+\frac{1}{2}\mbox{\boldmath $X$}^*_{\!(k+4)\bar{n}} 
  \left(\tan\!\theta_{\scriptscriptstyle W}\mbox{\boldmath $X$}^*_{\!1n}-\mbox{\boldmath $X$}^*_{\!2n}   \right)
  ({\cal D}^s_{\left(k+2\right)m}+i{\cal D}^s_{\left(k+7\right)m})  \,.
\end{align}
  
\bigskip\medskip
\noindent{\bf Charged Scalar-Neutral Lepton-Charged Lepton Vertices}
\[
{\cal L}=g_{\scriptscriptstyle 2}\overline{\Psi}(\chi^-_{\bar{n}})\left[
         \widetilde{\cal C}^{\chi L}_{\bar{n}mn}\frac{1-\gamma_5}{2}
        +\widetilde{\cal C}^{\chi R}_{\bar{n}mn}\frac{1+\gamma_5}{2}  \right]\Psi(\chi^0_n)\Phi(\phi^-_m) +\text{h.c.}
\]
where
\begin{align}
\widetilde{\cal C}^{\chi L}_{\bar{n}mn}=
&-\mbox{\boldmath $V$}^*_{1\bar{n}}\mbox{\boldmath $X$}^*_{\!3n}{\cal D}^l_{1m}
 +\frac{\sqrt{2}}{2}\mbox{\boldmath $V$}^*_{2\bar{n}}\left(-\tan\!\theta_{\scriptscriptstyle W}  
   \mbox{\boldmath $X$}_{\!1n}
  -\mbox{\boldmath $X$}_{\!2n}\right){\cal D}^l_{1m}\nonumber\\
&-\sqrt{2}\tan\!\theta_{\scriptscriptstyle W} \mbox{\boldmath $V$}^*_{\left(j+2\right)\bar{n}}
   \mbox{\boldmath $X$}_{\!1n} {\cal D}^l_{\left(j+5\right)m}
 -\frac{y_{e_j}}{g_{\scriptscriptstyle 2}}\left(
   \mbox{\boldmath $V$}^*_{\left(j+2\right)\bar{n}}\mbox{\boldmath $X$}_{\!4n} {\cal D}^l_{\left(j+2\right)m}    
  -\mbox{\boldmath $V$}^*_{\left(j+2\right)\bar{n}}\mbox{\boldmath $X$}_{\!(j+4)n}{\cal D}^l_{2m} \right)
              \nonumber\\
&-\frac{\lambda_{ijk}}{g_{\scriptscriptstyle 2}}\mbox{\boldmath $V$}^*_{\left(k+2\right)\bar{n}}
          \mbox{\boldmath $X$}_{\!(i+4)n} {\cal D}^l_{\left(j+2\right)m}    \nonumber  \\        
\widetilde{\cal C}^{\chi R}_{\bar{n}mn}=
&-\mbox{\boldmath $U$}^*_{1\bar{n}} \mbox{\boldmath $X$}^*_{\!4n} {\cal D}^l_{2m}
 -\mbox{\boldmath $U$}^*_{1\bar{n}} \mbox{\boldmath $X$}^*_{\!(k+4)n} {\cal D}^l_{\left(k+2\right)m}
 +\frac{\sqrt{2}}{2} \mbox{\boldmath $U$}^*_{2\bar{n}}\left( 
    \tan\!\theta_{\scriptscriptstyle W}\mbox{\boldmath $X$}^*_{\!1n}+\mbox{\boldmath $X$}^*_{\!2n}
    \right){\cal D}^l_{2m}\nonumber\\
&+\frac{\sqrt{2}}{2} \mbox{\boldmath $U$}^*_{\left(k+2\right)\bar{n}}\left( 
    \tan\!\theta_{\scriptscriptstyle W}\mbox{\boldmath $X$}^*_{\!1n} +\mbox{\boldmath $X$}^*_{\!2n}
    \right){\cal D}^l_{\left(k+2\right)m}           \nonumber\\
&-\frac{y_{e_k}}{g_{\scriptscriptstyle 2}}\left(\mbox{\boldmath $U$}^*_{\left(k+2\right)\bar{n}}
    \mbox{\boldmath $X$}^*_{\!4n} {\cal D}^l_{\left(k+5\right)m}   
         - \mbox{\boldmath $U$}^*_{2\bar{n}}\mbox{\boldmath $X$}^*_{\!(k+4)n} {\cal D}^l_{\left(k+5\right)m} \right)
             \nonumber\\
&-\frac{\lambda^*_{ijk}}{g_{\scriptscriptstyle 2}}\mbox{\boldmath $U$}^*_{\left(j+2\right)\bar{n}}
               \mbox{\boldmath $X$}^*_{\!(i+4)n} {\cal D}^l_{\left(k+5\right)m} \,.              
\end{align}  
 
\bigskip\medskip
\noindent{\bf Neutral Scalar-Squark-Squark Vertices: Down-Sector}
\[ {\cal L} =  \mbox{\boldmath $g$}^d_{abm} \,
 \Phi^{\dagger}({\tilde{d}_{a}})  \Phi({\tilde{d}_{b}})
 \Phi(\phi_{m}^0)
\]
where
\begin{align}
\mbox{\boldmath $g$}^d_{abm}=
&\frac{g_{\scriptscriptstyle 2} M_{\scriptscriptstyle Z}}{\cos\!{\theta_{\scriptscriptstyle W}}}
  \left(\frac{1}{2}-\frac{1}{3}\sin^2\!{\theta_{\scriptscriptstyle W}}\right)
  \left(\cos\!\beta \,{\cal D}^s_{2m}-\sin\!\beta \,{\cal D}^s_{1m}\right){\cal D}^{d*}_{qa} {\cal D}^d_{qb}
   \nonumber\\    
+&\frac{g_{\scriptscriptstyle 2} M_{\scriptscriptstyle Z}}{\cos\!{\theta_{\scriptscriptstyle W}}}
  \left(\frac{1}{3}\sin^2\!{\theta_{\scriptscriptstyle W}}\right)
  \left(\cos\!\beta \,{\cal D}^s_{2m}-\sin\!\beta \, {\cal D}^s_{1m} \right)
  {\cal D}^{d*}_{\left(q+3\right)a} {\cal D}^d_{\left(q+3\right)b} \nonumber\\ 
-&\sqrt{2}\,y_{d_q}\,m_{d_q} {\cal D}^{d*}_{qa} {\cal D}^d_{qb} {\cal D}^s_{2m} 
 -\sqrt{2}\,y_{d_q}\,m_{d_q} 
  {\cal D}^{d*}_{\left(q+3\right)a}{\cal D}^d_{\left(q+3\right)b}  {\cal D}^s_{2m} \nonumber\\  
+&\frac{1}{\sqrt{2}}\left(\mu^*_0 \,\delta_{pq} \,y_{d_q} +\mu^*_i\lambda^{'}_{ipq}\right)
   {\cal D}^{d*}_{\left(q+3\right)a}{\cal D}^d_{pb}  
   \left({\cal D}^s_{1m}+i{\cal D}^s_{6m}\right)  \nonumber\\
+&\frac{1}{\sqrt{2}}\left(\mu_0\, \delta_{pq}\, y_{d_q} +\mu_i\lambda^{'*}_{ipq}\right)
   {\cal D}^{d*}_{pa}{\cal D}^{d}_{\left(q+3\right)b}  
   \left({\cal D}^s_{1m}-i{\cal D}^s_{6m}\right) \nonumber\\
-&\frac{1}{\sqrt{2}}A^D_{pq} {\cal D}^{d*}_{\left(q+3\right)a}{\cal D}^d_{pb}   
  \left({\cal D}^s_{2m}+i{\cal D}^s_{7m}\right)
 -\frac{1}{\sqrt{2}}A^{D*}_{pq}{\cal D}^{d*}_{pa}   {\cal D}^d_{\left(q+3\right)b}
  \left({\cal D}^s_{2m}-i{\cal D}^s_{7m}\right) 
  \nonumber\\
-&\frac{1}{\sqrt{2}}A^{\lambda '}_{jpq} {\cal D}^{d*}_{\left(q+3\right)a}{\cal D}^d_{pb}  
  \left({\cal D}^s_{\left(j+2\right)m}+i{\cal D}^s_{\left(j+7\right)m}\right)\nonumber\\
-&\frac{1}{\sqrt{2}}A^{\lambda '*}_{jpq} {\cal D}^{d*}_{pa}  {\cal D}^{d}_{\left(q+3\right)b} 
  \left({\cal D}^s_{\left(j+2\right)m}-i{\cal D}^s_{\left(j+7\right)m}\right) \nonumber\\
-&\frac{1}{\sqrt{2}}
  \left[m_{d_p}\lambda^{'}_{ipq}{\cal D}^d_{\left(p+3\right)b} {\cal D}^{d*}_{\left(q+3\right)a} 
       +m_{d_q}\lambda^{'}_{ipq}  {\cal D}^{d*}_{qa} {\cal D}^d_{pb}  \right]
  \left({\cal D}^s_{\left(i+2\right)m}+i{\cal D}^s_{\left(i+7\right)m}\right)        \nonumber\\  
-&\frac{1}{\sqrt{2}}
  \left[m_{d_p}\lambda^{'*}_{ipq}{\cal D}^{d*}_{\left(p+3\right)a} {\cal D}^{d}_{\left(q+3\right)b}
       +m_{d_q}\lambda^{'*}_{ipq}  {\cal D}^{d}_{qb} {\cal D}^{d*}_{pa}\right]
  \left({\cal D}^s_{\left(i+2\right)m}-i{\cal D}^s_{\left(i+7\right)m}\right)   \,.  
\end{align}

\bigskip\medskip
\noindent{\bf Neutral Scalar-Squark-Squark Vertices: Up-Sector}
\[ {\cal L} =  \mbox{\boldmath $g$}^u_{abm} \,
 \Phi^{\dagger}({\tilde{u}_{a}})  \Phi({\tilde{u}_{b}})
 \Phi(\phi_{m}^0)
\]
where
\begin{align}
\mbox{\boldmath $g$}^u_{abm}
=&\frac{g_{\scriptscriptstyle 2} M_{\scriptscriptstyle Z}}{\cos\!{\theta_{\scriptscriptstyle W}}}
  \left(-\frac{1}{2}+\frac{2}{3}\sin^2\!{\theta_{\scriptscriptstyle W}}\right)
  \left(\cos\!\beta\, {\cal D}^s_{2m}-\sin\!\beta \,{\cal D}^s_{1m}\right){\cal D}^{u*}_{qa} {\cal D}^u_{qb} 
  \nonumber\\      
-&\frac{g_{\scriptscriptstyle 2} M_{\scriptscriptstyle Z}}{\cos\!{\theta_{\scriptscriptstyle W}}}
  \left(\frac{2}{3}\sin^2\!{\theta_{\scriptscriptstyle W}}\right)
  \left(\cos\!\beta\, {\cal D}^s_{2m} -\sin\!\beta \,{\cal D}^s_{1m}\right)
  {\cal D}^{u*}_{\left(q+3\right)a}{\cal D}^u_{\left(q+3\right)b}  \nonumber \\  
-&\sqrt{2}\,y_{u_l}m_{u_l}V_{\scriptscriptstyle C\!K\!M}^{lp*}
                          V_{\scriptscriptstyle C\!K\!M}^{lq} 
  {\cal D}^{u*}_{pa} {\cal D}^u_{qb}  {\cal D}^s_{1m} 
 -\sqrt{2}\,y_{u_q}m_{u_q}
  {\cal D}^{u*}_{\left(q+3\right)a}{\cal D}^u_{\left(q+3\right)b} {\cal D}^s_{1m}\nonumber\\    
+&\frac{1}{\sqrt{2}} y_{u_q}V_{\scriptscriptstyle C\!K\!M}^{qp} 
  \left[\mu^*_0\left({\cal D}^s_{2m}-i{\cal D}^s_{7m}\right) 
       +\mu^*_j\left({\cal D}^s_{\left(j+2\right)m}-i{\cal D}^s_{\left(j+7\right)m}\right)\right]
  {\cal D}^{u*}_{\left(q+3\right)a}{\cal D}^u_{pb}   \nonumber\\     
+&\frac{1}{\sqrt{2}} y_{u_q}V_{\scriptscriptstyle C\!K\!M}^{qp*}
  \left[\mu_0 \left({\cal D}^s_{2m}+i{\cal D}^s_{7m}\right)
       +\mu_j \left({\cal D}^s_{\left(j+2\right)m}+i{\cal D}^s_{\left(j+7\right)m}\right) \right] 
  {\cal D}^{u*}_{pa} {\cal D}^{u}_{\left(q+3\right)b}  \nonumber\\    
-&\frac{1}{\sqrt{2}}A^U_{pq} {\cal D}^{u*}_{\left(q+3\right)a}{\cal D}^u_{pb} 
  \left({\cal D}^s_{1m}-i{\cal D}^s_{6m}\right) \nonumber\\
-&\frac{1}{\sqrt{2}}A^{U*}_{pq} {\cal D}^{u*}_{pa} {\cal D}^{u}_{\left(q+3\right)b}
  \left({\cal D}^s_{1m}+i{\cal D}^s_{6m}\right) \,.
\end{align}

\bigskip\medskip
\noindent{\bf Cubic Neutral Scalar Vertices}
\[{\cal L} =  \mbox{\boldmath $g$}^0_{abm} \Phi(\phi_a^0)\Phi(\phi_b^0)\Phi(\phi_m^0)\]
where
\begin{align}
\mbox{\boldmath $g$}^0_{abm}=
&\frac{g_{\scriptscriptstyle 2} M_{\scriptscriptstyle Z}}{4\cos\theta_{\scriptscriptstyle W}}
   \left(\cos\!\beta {\cal D}^s_{2m}-\sin\!\beta {\cal D}^s_{1m}\right)\times \nonumber\\
&\qquad\qquad\left({\cal D}^s_{1a}{\cal D}^s_{1b} 
        +{\cal D}^s_{6a}{\cal D}^s_{6b}
      -{\cal D}^s_{2a}{\cal D}^s_{2b}
      -{\cal D}^s_{7a}{\cal D}^s_{7b}
      -{\cal D}^s_{\left(q+2\right)a}{\cal D}^s_{\left(q+2\right)b}
      -{\cal D}^s_{\left(q+7\right)a}{\cal D}^s_{\left(q+7\right)b}\right)\nonumber\\
+&\,\text{permutations in ($a$,$b$,$m$)}\,.                         
\end{align} 

\newpage
\noindent{\bf Neutral Scalar-Charged Scalar-Charged Scalar Vertices}
\[{\cal L} =  \mbox{\boldmath $g$}^-_{abm} \Phi^\dagger(\phi_a^-)\Phi(\phi_b^-)\Phi(\phi_m^0)\]
where
\begin{align}
\mbox{\boldmath $g$}^-_{abm}=
&-\frac{1}{2}\frac{g_{\scriptscriptstyle 2} M_{\scriptscriptstyle Z}}{\cos\theta_{\scriptscriptstyle W}} \sin\!\beta\,
        {\cal D}^s_{1m} {\cal D}^{l*}_{1a}{\cal D}^{l}_{1b}
 +\frac{g_{\scriptscriptstyle 2} M_{\scriptscriptstyle Z}}{\cos\theta_{\scriptscriptstyle W}}
  \left(-\frac{1}{2}+\sin^2\!\theta_{\scriptscriptstyle W}\right) \cos\!\beta\,
        {\cal D}^s_{2m} {\cal D}^{l*}_{1a}{\cal D}^{l}_{1b}  \nonumber\\ 
&-\frac{1}{2}\frac{g_{\scriptscriptstyle 2} M_{\scriptscriptstyle Z}}{\cos\theta_{\scriptscriptstyle W}}
  \left(1-\sin^2\!\theta_{\scriptscriptstyle W}\right)
  \left[\cos\!\beta\, \left({\cal D}^s_{1m} -i{\cal D}^s_{6m} \right)
       +\sin\!\beta\, \left({\cal D}^s_{2m} +i{\cal D}^s_{7m} \right) \right]
         {\cal D}^{l*}_{2a}{\cal D}^l_{1b}  \nonumber\\    
&-\frac{1}{2}\frac{g_{\scriptscriptstyle 2} M_{\scriptscriptstyle Z}}{\cos\theta_{\scriptscriptstyle W}}
  \left(1-\sin^2\!\theta_{\scriptscriptstyle W}\right)
  \left[\cos\!\beta\, \left({\cal D}^s_{1m} +i{\cal D}^s_{6m} \right)
        +\sin\!\beta\, \left({\cal D}^s_{2m} -i{\cal D}^s_{7m} \right)\right]
         {\cal D}^{l*}_{1a} {\cal D}^l_{2b}    \nonumber\\
&-\frac{1}{2}\frac{g_{\scriptscriptstyle 2} M_{\scriptscriptstyle Z}}{\cos\theta_{\scriptscriptstyle W}}
  \left(1-\sin^2\!\theta_{\scriptscriptstyle W}\right)
  \left(\sin\!\beta\, {\cal D}^l_{1b}+\cos\!\beta\, {\cal D}^l_{2b}\right) 
  \left({\cal D}^s_{\left(i+2\right)m}+i{\cal D}^s_{\left(i+7\right)m}\right){\cal D}^{l*}_{\left(q+2\right)a}   
  \nonumber\\    
&-\frac{1}{2}\frac{g_{\scriptscriptstyle 2} M_{\scriptscriptstyle Z}}{\cos\theta_{\scriptscriptstyle W}}
  \left(1-\sin^2\!\theta_{\scriptscriptstyle W}\right)
  \left(\sin\!\beta\,{\cal D}^{l*}_{1a} +\cos\!\beta\,{\cal D}^{l*}_{2a}\right)
  \left({\cal D}^s_{\left(i+2\right)m}-i{\cal D}^s_{\left(i+7\right)m}\right) {\cal D}^{l}_{\left(q+2\right)b}
  \nonumber\\                     
&-\frac{1}{2}\frac{g_{\scriptscriptstyle 2} M_{\scriptscriptstyle Z}}{\cos\theta_{\scriptscriptstyle W}}\cos\!\beta\,
        {\cal D}^s_{2m} {\cal D}^{l*}_{2a}  {\cal D}^l_{2b}
 -\frac{g_{\scriptscriptstyle 2} M_{\scriptscriptstyle Z}}{\cos\theta_{\scriptscriptstyle W}}
  \left(\frac{1}{2}-\sin^2\!\theta_{\scriptscriptstyle W}\right)\sin\!\beta\,
        {\cal D}^s_{1m} {\cal D}^{l*}_{2a}  {\cal D}^l_{2b}    
        \nonumber\\           
&+\frac{g_{\scriptscriptstyle 2} M_{\scriptscriptstyle Z}}{\cos\theta_{\scriptscriptstyle W}}
  \left(\frac{1}{2}-\sin^2\!\theta_{\scriptscriptstyle W}\right)
  \left(\cos\!\beta\, {\cal D}^s_{2m}-\sin\!\beta\, {\cal D}^s_{1m} \right)
   {\cal D}^{l*}_{\left(q+2\right)a}{\cal D}^{l}_{\left(q+2\right)b} \nonumber\\           
&+\frac{g_{\scriptscriptstyle 2} M_{\scriptscriptstyle Z}}{\cos\theta_{\scriptscriptstyle W}}
  \left(\sin^2\!\theta_{\scriptscriptstyle W}\right)
  \left(\cos\!\beta\, {\cal D}^s_{2m}-\sin\!\beta\, {\cal D}^s_{1m}\right) 
  {\cal D}^{l*}_{\left(q+5\right)a}{\cal D}^{l}_{\left(q+5\right)b}  \nonumber\\  
&-\sqrt{2}\,y_{e_q} m_{e_q}
   {\cal D}^{l*}_{\left(q+2\right)a}{\cal D}^{l}_{\left(q+2\right)b}{\cal D}^s_{2m} 
 -\sqrt{2}\,y_{e_q} m_{e_q}  {\cal D}^{l*}_{\left(q+5\right)a}  {\cal D}^l_{\left(q+5\right)b}
  {\cal D}^s_{2m}  \nonumber\\      
&-\frac{1}{\sqrt{2}}\mu^*_q y_{e_q} {\cal D}^{l*}_{\left(q+5\right)a}{\cal D}^l_{2b} 
  \left({\cal D}^s_{1m}+i{\cal D}^s_{6m}\right)
 -\frac{1}{\sqrt{2}}\mu_q y_{e_q}{\cal D}^{l*}_{2a} {\cal D}^{l}_{\left(q+5\right)b}
  \left({\cal D}^s_{1m}-i{\cal D}^s_{6m}\right) \nonumber\\  
&-\frac{1}{\sqrt{2}}\mu^*_q y_{e_q} {\cal D}^{l*}_{\left(q+5\right)a}{\cal D}^l_{1b} 
  \left({\cal D}^s_{2m}+i{\cal D}^s_{7m}\right) 
 -\frac{1}{\sqrt{2}}\mu_q y_{e_q} {\cal D}^{l*}_{1a}{\cal D}^{l}_{\left(q+5\right)b}
  \left({\cal D}^s_{2m}-i{\cal D}^s_{7m}\right) 
   \nonumber\\          
&+\frac{1}{\sqrt{2}} \left(\mu^*_0 \delta_{pq}y_{e_q}+\mu^*_i\lambda_{ipq}\right)
  {\cal D}^{l*}_{\left(q+5\right)a}{\cal D}^{l}_{\left(p+2\right)b}     
  \left({\cal D}^s_{1m}+i{\cal D}^s_{6m}\right) \nonumber\\                  
&+\frac{1}{\sqrt{2}}\left(\mu_0 \delta_{pq} y_{e_q}+\mu_i\lambda^*_{ipq}\right) 
  {\cal D}^{l*}_{\left(p+2\right)a} {\cal D}^{l}_{\left(q+5\right)b}
  \left({\cal D}^s_{1m}-i{\cal D}^s_{6m}\right) \nonumber\\ 
&+\frac{1}{\sqrt{2}}\left(\mu^*_0\delta_{pq} y_{e_q}+\mu^*_i\lambda_{ipq}\right)
  {\cal D}^{l*}_{\left(q+5\right)a}{\cal D}^l_{1b} 
  \left({\cal D}^s_{\left(p+2\right)m}+i{\cal D}^s_{\left(p+7\right)m}\right) 
  \nonumber\\           
&+\frac{1}{\sqrt{2}}\left(\mu_0\delta_{pq}y_{e_q}+\mu_i\lambda^*_{ipq}\right)
  {\cal D}^{l*}_{1a}{\cal D}^{l}_{\left(q+5\right)b}
  \left({\cal D}^s_{\left(p+2\right)m}-i{\cal D}^s_{\left(p+7\right)m}\right) \nonumber\\
&+\frac{1}{\sqrt{2}}m_{e_q}y_{e_q} {\cal D}^{l*}_{\left(q+2\right)a}{\cal D}^l_{2b}
   \left({\cal D}^s_{\left(q+2\right)m}+i{\cal D}^s_{\left(q+7\right)m}\right)
   \nonumber\\                
&+\frac{1}{\sqrt{2}}m_{e_q}y_{e_q} {\cal D}^{l*}_{2a}{\cal D}^{l}_{\left(q+2\right)b}
   \left({\cal D}^s_{\left(q+2\right)m}-i{\cal D}^s_{\left(q+7\right)m}\right)
   \nonumber                    
\end{align}  
\newpage
\begin{align}    
&-\frac{1}{\sqrt{2}}m_{e_p} \lambda_{jpq}{\cal D}^{l*}_{\left(q+5\right)a} {\cal D}^{l}_{\left(p+5\right)b} 
  \left({\cal D}^s_{\left(j+2\right)m}+i{\cal D}^s_{\left(j+7\right)m}\right)  \nonumber\\     
&-\frac{1}{\sqrt{2}}m_{e_p} \lambda^*_{jpq}{\cal D}^{l*}_{\left(p+5\right)a}{\cal D}^{l}_{\left(q+5\right)b}
  \left({\cal D}^s_{\left(j+2\right)m}-i{\cal D}^s_{\left(j+7\right)m}\right) 
  \nonumber\\
&-\frac{1}{\sqrt{2}}m_{e_q}\lambda_{jpq}{\cal D}^{l*}_{\left(q+2\right)a}{\cal D}^{l}_{\left(p+2\right)b}
  \left({\cal D}^s_{\left(j+2\right)m}+i{\cal D}^s_{\left(j+7\right)m}\right) 
  \nonumber\\         
&-\frac{1}{\sqrt{2}}m_{e_q}\lambda^*_{jpq} 
  {\cal D}^{l*}_{\left(p+2\right)a}{\cal D}^{l}_{\left(q+2\right)b}
  \left({\cal D}^s_{\left(j+2\right)m}-i{\cal D}^s_{\left(j+7\right)m}\right)\nonumber \\     
&-\frac{1}{\sqrt{2}}A^E_{pq} 
  {\cal D}^{l*}_{\left(q+5\right)a} {\cal D}^l_{\left(p+2\right)b}
  \left({\cal D}^s_{2m}+i{\cal D}^s_{7m}\right) 
 -\frac{1}{\sqrt{2}}A^{E*}_{pq} 
  {\cal D}^{l*}_{\left(p+2\right)a} {\cal D}^l_{\left(q+5\right)b} 
  \left({\cal D}^s_{2m}-i{\cal D}^s_{7m}\right)      \nonumber\\   
&+\frac{1}{\sqrt{2}}A^E_{pq}                    
  {\cal D}^{l*}_{\left(q+5\right)a} {\cal D}^l_{2b} 
  \left({\cal D}^s_{\left(p+2\right)m}+i{\cal D}^s_{\left(p+7\right)m}\right)   
 +\frac{1}{\sqrt{2}}A^{E*}_{pq}
  {\cal D}^{l*}_{2a} {\cal D}^{l}_{\left(q+5\right)b}
  \left({\cal D}^s_{\left(p+2\right)m}-i{\cal D}^s_{\left(p+7\right)m}\right)   
  \nonumber\\   
&-\frac{1}{\sqrt{2}}  A^\lambda_{jpq} 
  {\cal D}^{l*}_{\left(q+5\right)a}{\cal D}^{l}_{\left(p+2\right)b}
  \left({\cal D}^s_{\left(j+2\right)m}+i{\cal D}^s_{\left(j+7\right)m}\right)  \nonumber\\                          
&-\frac{1}{\sqrt{2}}  A^{\lambda*}_{jpq}
  {\cal D}^{l*}_{\left(p+2\right)a} {\cal D}^l_{\left(q+5\right)b} 
  \left({\cal D}^s_{\left(j+2\right)m}-i{\cal D}^s_{\left(j+7\right)m}\right) \,.                                
\end{align}

\newpage
\begin{figure}[h]
\begin{minipage}[]{0.327\linewidth} 
\includegraphics[scale=1]{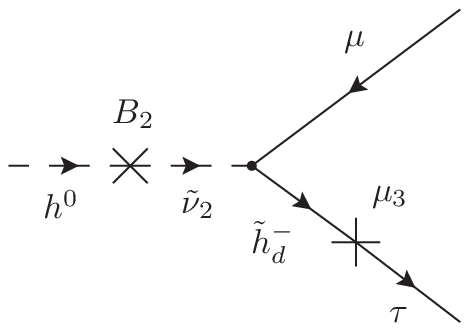}
\end{minipage}
\begin{minipage}[]{0.327\linewidth} 
\includegraphics[scale=1]{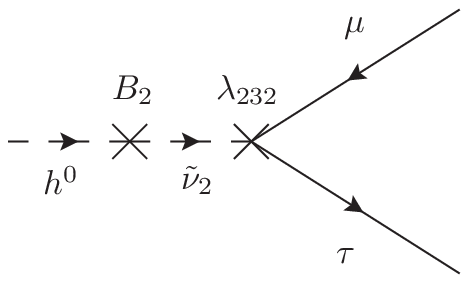}
\end{minipage}
\begin{minipage}[]{0.327\linewidth} 
\includegraphics[scale=1]{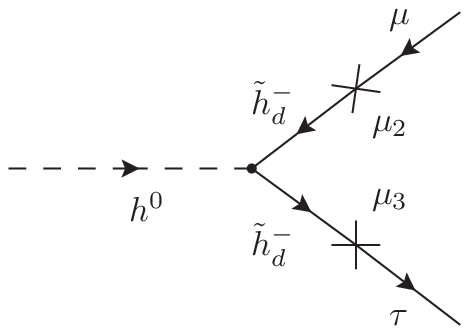}
\end{minipage}
\end{figure}
\noindent 
Fig. 1. Left panel: An example of $B_2 \mu_3$ contribution to $h^0 \rightarrow \tau^-\mu^+$ via tree diagram. The Higgsino $\tilde{h}^-_d$ transforms into charged lepton $\tau$ via RPV parameter $\mu_3$, while the light Higgs transforms into sneutrino $\tilde{\nu}_2$ via $B_2$. \\
Middle panel: An example of $B_2 \lambda_{232}$ contribution to $h^0 \rightarrow \tau^-\mu^+$ via tree diagram. The light Higgs transforms into sneutrino $\tilde{\nu}_2$ via RPV parameter $B_2$ and then couples to $\mu$ and $\tau$ via trilinear RPV parameter $\lambda_{232}$. \\
Right panel: An example of $\mu_2 \mu_3$ contribution to $h^0 \rightarrow \tau^-\mu^+$ via tree diagram. The Higgsino $\tilde{h}^-_d$ mixes with charged leptons $\mu$ and $\tau$ via RPV parameters $\mu_2$ and $\mu_3$ separately. \\

\begin{figure}[h]
\begin{minipage}[]{0.495\linewidth} 
\includegraphics[scale=1.05]{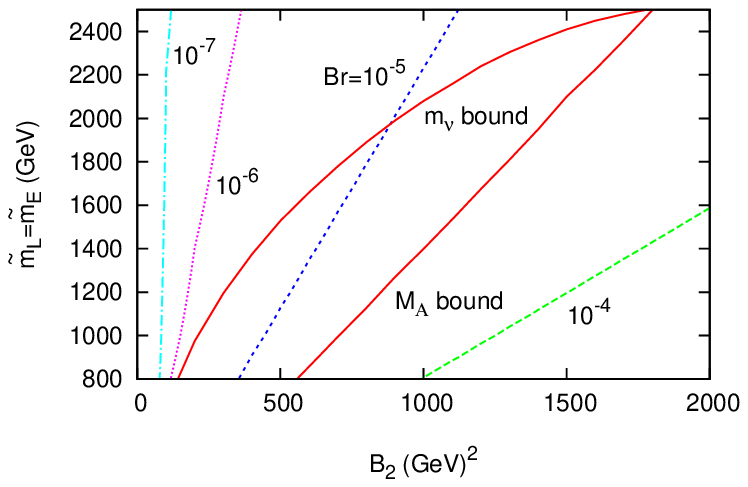}
\end{minipage}
\begin{minipage}[]{0.495\linewidth} 
\includegraphics[scale=1.05]{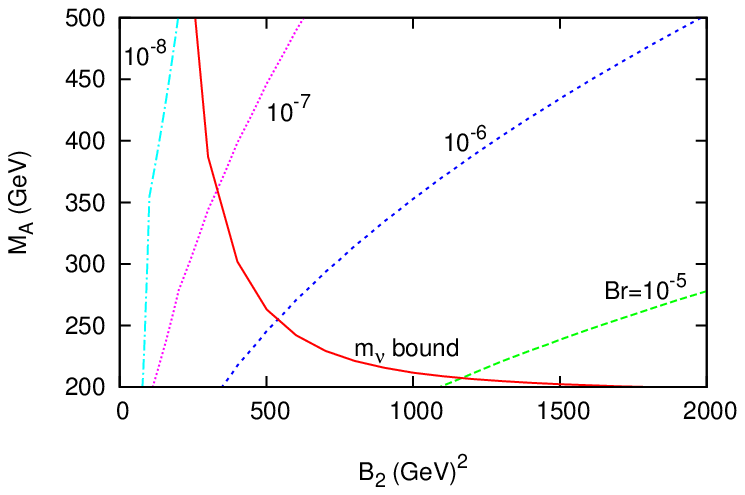}
\end{minipage}
\end{figure} 
\noindent
Fig. 2. Left panel: Branching ratio from $B_2\,\lambda_{232}$, with $M_2=2500$ GeV, $\mu_0=1800$ GeV$=A_u=-A_d$\,, $\tan\beta=60$. $\lambda_{232}$ is set to be the maximum dependent on the values of $B_2$ and $\tilde{m}^2_L=\tilde{m}^2_E$\,. The Solid red line ($m_\nu$ bound) comes from demanding that the 22 element of the neutrino mass matrix $<$ 1 eV, while the right-hand side of the $M_A$ bound line is the area with CP-odd neutral Higgs mass $M_A<$ 200 GeV. \\
Right panel: Branching ratio from $B_2\,\lambda_{232}$, with $M_2=2500$ GeV, $\tilde{m}^2_{Lii}=\tilde{m}^2_{Eii}$=$(2500 ~\text{GeV})^2$, $\mu_0=1800$ GeV$=A_u=-A_d$\,, $\tan\beta=60$, $\lambda_{232}=1.7488$. The Solid red line ($m_\nu$ bound) comes from demanding that the 22 element of the neutrino mass matrix $<$ 1 eV.

\begin{figure}[h]
\begin{minipage}[]{0.495\linewidth} 
\includegraphics[scale=1.05]{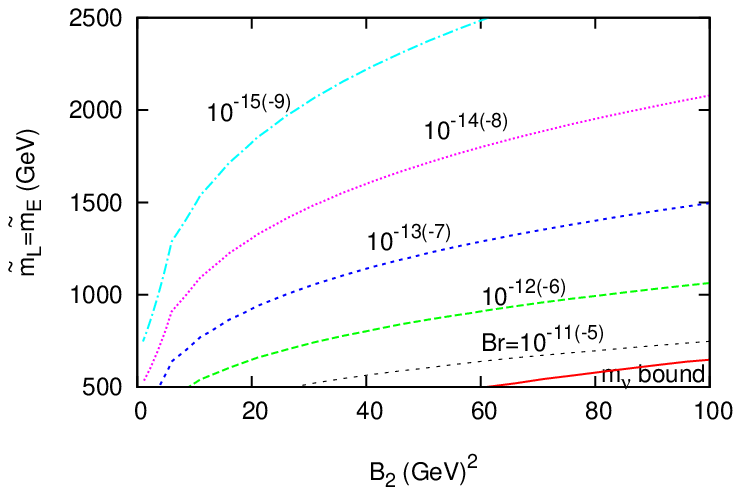}
\end{minipage}
\begin{minipage}[]{0.495\linewidth} 
\includegraphics[scale=1.05]{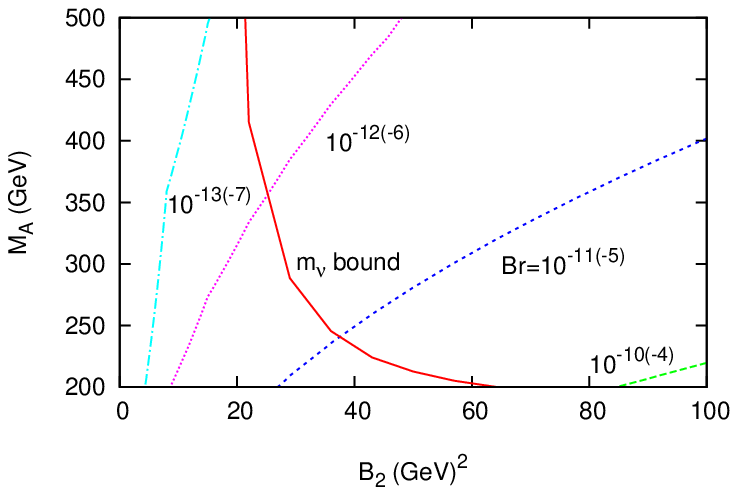}
\end{minipage}
\end{figure} 
\noindent
Fig. 3. Left panel: Branching ratio from $B_2\,A^\lambda_{232}$, with $M_2=2500$ GeV, $M_A\cong$ 200 to 202 GeV, $\mu_0=1800 ~\text{GeV}=A_u=-A_d$\,, $\tan\beta=60$, $A^\lambda_{232}=2500$ GeV(TeV). The Solid red line ($m_\nu$ bound) comes from demanding that the 22 element of the neutrino mass matrix $<$ 1 eV.\\
Right panel: Branching ratio from $B_2\,A^\lambda_{232}$, with $M_2=2500~\text{GeV}$, $\tilde{m}^2_{Lii}=\tilde{m}^2_{Eii}=(500 ~\text{GeV})^2$, $\mu_0=1800~\text{GeV}=A_u=-A_d$\,, $\tan\beta=60$, $A^\lambda_{232}=2500$ GeV(TeV). The Solid red line ($m_\nu$ bound) comes from demanding that the 22 element of the neutrino mass matrix $<$ 1 eV.
\end{document}